\documentclass[journal]{IEEEtran}
\usepackage{tikz}
\usepackage{pgfplots}
\pgfplotsset{compat=1.17}
\usetikzlibrary{external}
\usetikzlibrary{patterns}
\usepackage{array}
\usepackage{lipsum,graphicx,multicol,multirow}
\usetikzlibrary{shapes,arrows,positioning,calc,decorations.pathreplacing}
\usetikzlibrary {arrows.meta,shapes.misc,calc}
\usepackage{amsmath, amssymb,amsfonts,amscd,xpatch}
\usepackage{cases}
\usepackage{mwe,comment}
\usepackage{xcolor,colortbl} % for table color
\usepackage[noadjust]{cite}

\ifCLASSOPTIONcompsoc
 \usepackage[caption=false,font=normalsize,labelfont=sf,textfont=sf]{subfig}
\else
 \usepackage[caption=false,font=footnotesize]{subfig}
\fi

\newcommand{\Energies}{\boldsymbol{E}}
\newcommand{\Energy}{E}
\newcommand{\AcEnergy}{E^{\text{acc}}} 
\newcommand{\Exp}[1]{\mathbb{E}\left[#1\right]} 
\newcommand{\Var}[1]{\mathrm{Var}\left[#1\right]}
\newcommand{\AvgExp}[1]{\overline{\mathbb{E}}\left[#1\right]} 
\newcommand{\AvgVar}[1]{\overline{\mathrm{Var}}\left[#1\right]}
\newcommand{\TExp}[1]{\mu_{\scriptscriptstyle #1}}
\newcommand{\TVar}[1]{\sigma^2_{\scriptscriptstyle #1}}
\newcommand{\Cov}[1]{\mathrm{Cov}\left[#1\right]} 
\newcommand{\Prob}{\mathbb{P}} 
 
\newcommand{\CrS}{\rho}
\newcommand{\AC}{\overline{R}_{\Energy}(\tau)}
\newcommand{\TAC}{\hat{R}_{\Energy}(\tau)}

\newcommand{\WE}{G^W}

\newcommand{\SNRe}{\mathrm{SNR}_{\mathrm{eff}}}

\newcounter{lemma}
\newtheorem{lemma}{Lemma}
\newtheorem{theorem}[lemma]{Theorem}
\newtheorem{corollary}[lemma]{Corollary}
\newtheorem{example}{Example}
\newtheorem{definition}{Definition}

\begin{document}
%%%%% paper title  %%%%%%%%%%%%%%%%%%%%%%%%%
\title{Temporal Energy Analysis of Symbol Sequences for Fiber Nonlinear Interference Modelling via Energy Dispersion Index}

\author{Kaiquan~Wu,~\IEEEmembership{Student Member, IEEE}, Gabriele~Liga,~\IEEEmembership{Member, IEEE}, Alireza~Sheikh,~\IEEEmembership{Member, IEEE}, Frans~M.~J.~Willems,~\IEEEmembership{Life Fellow, IEEE}, and Alex~Alvarado,~\IEEEmembership{Senior Member, IEEE}% <-this % stops a space
\thanks{This work is supported by the Netherlands Organisation for Scientific Research via the VIDI Grant ICONIC (project number 15685). The work of Alex Alvarado is supported by the European Research Council (ERC) under the European Union’s Horizon 2020 research and innovation programme (grant agreement No. 757791). The work of G.~Liga is supported by the EuroTechPostdoc programme under the European Union’s Horizon 2020 research and innovation programme (Marie Skłodowska-Curie grant agreement No 754462).}
\thanks{The authors are with the Information and Communication Theory Lab, Signal Processing Systems Group, Department of Electrical Engineering, Eindhoven University of Technology, Eindhoven 5600 MB, The Netherlands (e-mails: \{k.wu, g.liga, a.sheikh, f.m.j.willems, a.alvarado\}@tue.nl).}
\thanks{A. Sheikh is with imec, Holst Centre, High Tech Campus 31, 5656 AE Eindhoven, The Netherlands (email: alireza.sheikh@imec.nl).}}% <-this % stops a space

%%%%%  The paper headers  %%%%%%%%%%%%%%%%%%%%%%%%
\markboth{Preprint, \today}%
{Shell \MakeLowercase{\textit{et al.}}: Bare Demo of IEEEtran.cls for IEEE Journals}

\maketitle

%%%% Abstract %%%%%%%%%%%%%%%%%%%%%%%%%%%%%%%
\begin{abstract}
The stationary statistical properties of independent, identically distributed (i.i.d.) input symbols provide insights on the induced nonlinear interference (NLI) during fiber transmission. For example, \emph{kurtosis} is known to predict the modulation format-dependent NLI. These statistical properties can be used in the design of probabilistic amplitude shaping (PAS), which is a popular scheme that relies on an amplitude shaper for increasing spectral efficiencies of fiber-optic systems. One property of certain shapers used in PAS---including constant-composition distribution matchers---that is often overlooked is that a time-dependency between amplitudes is introduced. This dependency results in symbols that are \emph{non}-i.i.d., which have time-varying statistical properties. Somewhat surprisingly, the effective signal-to-noise ratio (SNR) in PAS has been shown to increase when the shaping blocklength decreases. This blocklength dependency of SNR has been attributed to time-varying statistical properties of the symbol sequences, in particular, to variation of the symbol energies. In this paper, we investigate the temporal energy behavior of symbol sequences, and introduce a new metric called \emph{energy dispersion index} (EDI). EDI captures the time-varying statistical properties of symbol energies. Numerical results show strong correlations between EDI and effective SNR, with absolute correlation coefficients above $99\%$ for different transmission distances.
\end{abstract}


\IEEEpeerreviewmaketitle

%%%%  Introduction  %%%%%%%%%%%%%%%%%%%%%%%%%%%%%
\section{Introduction}\label{sec:intro}

%%%%%% Paragraph 1
% A high-level introduction to shaping and coding in fiber-optic communications. Pointing out NLI in the nonlinear fiber remains a significant limitation for improving performance.
\IEEEPARstart{C}{onstellation} shaping (CS) and forward error correction (FEC) are two crucial elements to realize near capacity-achieving transmission for the additive white Gaussian noise (AWGN) channel. In recent years, a novel coded modulation framework called probabilistic amplitude shaping (PAS) \cite{bocherer2015bandwidth} has attracted wide attention for its capacity-achieving performance. PAS elegantly integrates FEC and probabilistic shaping (PS) \cite{cho2019probabilistic}, where the PS functionality is enabled by an amplitude shaper. One popular amplitude shaper is constant-composition distribution matcher (CCDM) \cite{schulte2015constant}. Other well-known shapers include multiset-partition distribution matcher (MPDM) \cite{fehenberger2018multiset}, product distribution matcher (PDM) \cite{steiner2018approaching}, enumerative sphere shaping (ESS) \cite{willems1993pragmatic,gultekin2017constellation}, and shell mapping \cite{khandani1993shaping}.

%%%%%% Paragraph 2
% Pointing out NLI in the nonlinear fiber remains a significant limitation for improving performance.
In the context of fiber optical communications, numerous studies about PAS have been carried out in simulation \cite{cho2016low, fehenberger2016probabilistic,amari2019introducing} and experiments \cite{buchali2016rate,ghazisaeidi2017advanced}. Record spectral efficiencies (SEs) have been achieved in field trials \cite{cho2018trans,olsson2018record}. However, unlike the AWGN channel, SEs in the nonlinear fiber channel are limited by the Kerr effect. This effect causes nonlinear interference (NLI), which becomes a substantial part of the total noise experienced by the transmitted signals \cite{agrawal2013nonlinear}. 

%%%%%% Paragraph 3-4
% Explain that the NLI is mainly determined by the time-averaged and time-varying statistics of the symbol sequences. Introduce studies on optimizing these two kinds of statistics.
Mitigation of NLI can be achieved by optimizing statistical properties of the transmitted symbol sequences. Given a number of constellation points, it is possible to change the constellation geometry \cite{qu2017geometrically,gumucs2020end} or the probability mass function (PMF) of the constellation symbols \cite{renner2017experimental,cho2016low,sillekens2018simple}. These techniques are often referred to as geometric and probabilistic shaping, respectively. In both cases, by assuming symbols to be independent identically distributed (i.i.d.), the stationary statistical properties of symbols are optimized. In the former, the support of the PMF is optimized, while in the later, the probabilities are optimized.

An alternative approach is to manipulate the \emph{temporal} structure of the symbol sequence, which is believed to exert great influence on the NLI \cite{agrell2014capacity,dar2014shaping}. The NLI can be modeled as inter-symbol interference (ISI) \cite{dar2014time}, i.e., the NLI experienced by a transmitted symbol only depends on adjacent symbols. The number of interfering symbols is usually referred to as channel memory. Therefore, transmitted symbols that comply to certain temporal structures could be used to suppress NLI. One example of this approach is the so-called temporal shaping introduced in \cite{dar2014shaping}, which generates correlated symbols from a ball-shaped constellation. By taking the correlation between symbol energies into account, \cite{dar2014shaping} showed that transmitting correlated symbols leads to a new correlated term compared to i.i.d. symbol sequences (see \cite[Eqs. (5) and (9)]{dar2014shaping}). Temporal shaping was later realized using a finite state machine \cite{yankov2017temporal}. Although improved tolerance to nonlinearities can be achieved by controlling the temporal structure of the symbol sequences, a guiding principle for the optimization of the temporal structure is still unknown. This could be due to the fact that the statistical analysis of non-i.i.d. symbol sequences is in general more difficult than its i.i.d. counterpart.

%% multi-dimensional constellation
% Introducing dependencies in multi-dimensional modulation formats improves tolerance to nonlinearities more significantly with respect to 2D modulation formats. More specifically, four-dimensional (4D) modulation formats provide nonlinear shaping gains by introducing dependencies across quadratures (I/Q) and polarization states (X/Y) \cite{chen2020analysis,van201911}. Further extending the dependency to the time dimension leads to 8D modulation formats. Higher nonlinear shaping gains were recently demonstrated in \cite{chen2019eight}. %Subsystems such as FEC and shapers in the PAS controlling the temporal behavior of the symbol sequences can also be modified. 

%%%%%% Paragraph 5
% Introduce studies of temporal structure and the NLI. 
PAS generates dependent symbols because the employed shaper imposes hidden temporal structure on the amplitude blocks. Recently, several studies based on the PAS architecture have observed the impact of the resulting temporal structure on the effective SNR. The effective SNR has been shown to depend on the shaping blocklength: using short shaping blocklengths has been shown to offer significant effective SNR gains due to a weaker presence of nonlinearities \cite{amari2019introducing,fehenberger2019analysis,goossens2019first,fehenberger2020mitigating}. In \cite{civelli2020interplay}, the relationship between carrier phase recovery and shaping blocklength on the nonlinear shaping gain was studied. In addition, the symbol mapping strategy, i.e., the way the shaped amplitudes are mapped to multi-dimensional symbols, has also been shown to be a crucial factor in the effective SNR performance of the system. It was observed in \cite{fehenberger2019analysis,skvortcov2020huffman} that instead of using four amplitude shapers independently for I/Q and X/Y dimensions, using one amplitude shaper across four dimensions improves the effective SNR.

%%%%%% Paragraph 6
% Under i.i.d. assumption, show the GN and the EGN models with metrics that is used to suppress NLI. Introduce the usage if metric in designing NLI-tolerant signaling scheme.
The assumption of i.i.d. input symbols can be justified in scenarios such as (i) uncoded transmission, and (ii) systems employing a random interleaver that breaks dependencies between symbols. Based on the i.i.d. assumption, state-of-the-art Gaussian noise (GN) and enhanced Gaussian noise (EGN) models provide insights about the connection between the stationary statistical properties of the transmitted signal and the NLI. The GN model concludes that the NLI power scales as the cube of the average symbol energy \cite{poggiolini2011analytical,poggiolini2013gn,carena2012modeling}. By relaxing the \emph{Gaussianity} assumption of the GN model \cite{dar2013properties}, the EGN model shows that the standardized fourth moment (also known as kurtosis\footnote{The sixth-order moment reflects self-channel interference, which is marginal in long distance transmission using wavelength-division multiplexing (WDM).}) of the transmitted symbols is an important metric in predicting the modulation-dependent NLI \cite{dar2014inter,carena2014accuracy,poggiolini2015simple}. To mitigate the NLI, an optimized PMF was designed in \cite{sillekens2018simple} with the help of kurtosis. In \cite{fehenberger2019analysisECOC}, the blockwise composition of the symbol sequences having low kurtosis was selected for transmission. 

%%%%%% Paragraph 7
% Under non-i.i.d. assumption with correlated symbols, show how the NLI is modeled.
To address the scenario of non-i.i.d. input symbols, the heuristic finite-memory GN model was proposed in \cite{agrell2014capacity}, which considers a time-windowed symbol energy within the channel memory for the computation of the NLI. Recently, to analyze the temporal structure of symbol sequences, it was shown in \cite{fehenberger2019analysis,fehenberger2020impact} that that symbol sequences with fewer clusters of identical symbols offer reduced NLI. A heuristic metric called run ratio was also proposed in \cite{fehenberger2019analysis}. The perturbative time-domain model of \cite{mecozzi2012nonlinear} can in principle also be used to analyze non-i.i.d. input symbols.

%%%%%% Paragraph 8-9
% Show the motivation and contributions of the paper, 1). Analyze constant-composition sequences are not i.i.d. 2). Propose Energy dispersion index.
The discussion above leaves multiple open questions about the effects of the temporal structure of the symbol sequences on the NLI. Perhaps the most important question is that a metric that accounts for time-varying statistical properties of symbol sequences and enables a precise NLI prediction is still unknown. This paper is devoted to presenting a detailed explanation for the blocklength dependency on effective SNR, and introducing a metric that is able to capture the effect of time-varying statistics on the NLI. 

The contributions of this paper are three. First, we show that the temporal structure of the symbol sequence shaped by CCDM yields correlated symbols. The second contribution is to introduce a novel metric called energy dispersion index (EDI) to evaluate the variance of the symbol energies within a window. This metric is inspired by the channel models in \cite{dar2014shaping,agrell2014capacity} and is a function of the autocorrelation function of the symbol energy's variance within a window. An almost perfect correlation between the EDI and the effective SNR for different transmission distances is observed in numerical analysis using CCDM. Lastly, we also give analytical expressions for the autocorrelation and the EDI of the QAM symbol sequences generated by CCDM. %Applications of EDI on other shaping algorithms will be left for future work.

%%%%%% Paragraph 10
% Show the organization of the paper.
The rest of the paper is organized as follows. A review of the channel memory estimation and relevant channel models for the optical fiber channel is presented in Sec.~\ref{sec:NLIMod}. In Sec.~\ref{sec:CCQAM}, the statistical properties of symbol sequences shaped by CCDM are studied. The EDI, which is the main contribution of the paper, is presented along with the simulation results in Sec.~\ref{sec:EDI} and Sec.~\ref{sec:NumAna}. Conclusions are drawn in Sec.~\ref{sec:Conc}.

%%%%  Sec  %%%%%%%%%%%%%%%%%%%%%%%%%%%%%%%%%%%
\section{Modeling NLI with Windowed Energy}\label{sec:NLIMod}
In this section, the estimation of memory in the fiber-optic channel and the effective SNR considering the NLI are reviewed. We then discuss the effects of temporal energy behavior of symbol sequences on the NLI, from the perspectives of channel models with memory \cite{mecozzi2012nonlinear,dar2014shaping,agrell2014capacity}. We end this section with a review of metrics available in the literature that predict the NLI induced by input symbols.

%%%% Subsec %%%%%%%%%%%%%%%%%%%%%%%%%
\subsection{Channel Memory and Effective SNR}\label{subsec:NLIMod}
In a nonlinear fiber channel, amplifiers along with conventional linear digital signal processing (DSP) effectively compensate chromatic dispersion (CD) and attenuation. In this paper, we consider the channel output after DSP steps excluding nonlinearity compensation. Thus, the residual noise comes mainly from amplified spontaneous emission (ASE) and NLI. The ASE noise is a random process independent of the signal, whereas the NLI noise is dependent on the signal. The interplay between CD and fiber nonlinearity induces nonlinear interactions among a number of past and future symbols. The number of interfering symbols is often referred to as the channel memory. 

For a fiber-optic system with group velocity dispersion $\beta_{2}$ and optical bandwidth $\Delta\omega$, the dispersive length $L_{\mathrm{D}}$ is defined as \cite[Sec.~II-B]{agrell2014capacity}
\begin{equation}\label{ChnMm1}
    L_{\mathrm{D}}=1 /\left(\Delta \omega^{2}\left|\beta_{2}\right|\right),
\end{equation}
where $L_{\mathrm{D}}$ indicates the distance scale over which pulse broadening effects become significant. Given the propagation distance $L$, the \emph{one-sided} channel memory $M$ is roughly approximated as \cite[Sec.~II-B]{agrell2014capacity}

\begin{equation}\label{ChnMm2}
    M \approx L/L_{\mathrm{D}}.
\end{equation}

At time instant $i$, given the transmitted symbol $X_i$, a vector consisting of $X_{i}$ along with its $M$ neighbouring symbols is denoted as $\boldsymbol{X}_{i-M}^{i+M} = [X_{i-M},\ldots,X_{i},X_{i+1},\ldots,X_{i+M}]$\footnote{\emph{Notation}: Throughout this paper, random variables are denoted by uppercase letters $X$ and their (deterministic) outcomes by the same letter in lowercase $x$. Sequences are denoted by boldface letters $\boldsymbol{X}$ (random) or $\boldsymbol{x}$ (deterministic). We use subscript and superscript to denote the boundary of a random sequences, i.e., $\boldsymbol{X}_{i}^{j}=[X_{i},X_{i+1},\ldots,X_{j}]$. %The Euclidean norm is denoted by $\|\cdot\|$. 
Expectations, variances and autocovariances are denoted by $\Exp{\cdot}$, $\Var{\cdot}$, and $\Cov{\cdot}$, respectively. %The sample mean and variance of a random sequence $\boldsymbol{X}$ are denoted by $\TExp{X}$ and $\TVar{X}$, respectively. 
The probability of $X_i=x$ is denoted by $\Prob_{X_i}(x)$. Conditional probability is denoted as $\Prob_{Y_i\mid X_i}(y|x)$. The imaginary unit is $\jmath \triangleq \sqrt{-1}$.}. By considering the NLI and the ASE as AWGNs, the received symbol $Y_i$ can be expressed as
\begin{equation}\label{yNLI}
    Y_{i} = X_{i} + Z_{\text{ASE},i}+Z_{\text{NLI},i}(\boldsymbol{X}_{i-M}^{i+M}).
\end{equation}
Due to the nonlinear interference, the NLI term $Z_{\text{NLI},i}$ is expressed as a function of $\boldsymbol{X}_{i-M}^{i+M}$. 

The SNR is usually evaluated under the implicit assumption of ergodic signal and noise processes. Therefore, the empirical estimations of the signal power and noise variance converge to the corresponding statistical values. Hence, the \emph{effective} SNR including the NLI is defined as
\begin{align}
        \SNRe & \triangleq \frac{\Exp{|X|^2}}{\Var{Z_{\text{ASE}}}+\Var{Z_{\text{NLI}}}} \label{effSNR0} \\
        & =
        \frac{\Exp{|X|^2}}{\Exp{|Y-X|^2}} {\:\approx\:} \frac{\TExp{|X|^2}}{\TExp{|Y-X|^2}}.\label{effSNR}
\end{align}
Often the r.h.s. of \eqref{effSNR} is used to approximate the effective SNR in simulations and experiments.

%%%% Subsec %%%%%%%%%%%%%%%%%%%%%%%%%
\subsection{Fiber Channel Models with Finite Memory}
We review related channel models that consider the temporal effects of the transmitted symbols and also fit the general model in \eqref{yNLI}. A first-order perturbation model can be derived from the nonlinear Schr\"{o}dinger equation in discrete-time domain \cite{mecozzi2012nonlinear}. Assuming a single channel at time instant $i=0$, $Z_{\text{NLI},0}$ is modeled as \cite[Eq.~(57)]{mecozzi2012nonlinear}
\begin{equation}\label{KnMd0}
   Z_{\text{NLI},0}(\boldsymbol{X}_{-\infty}^{+\infty})= \jmath \gamma \sum_{h=-\infty}^{+\infty} \sum_{j=-\infty}^{+\infty} \sum_{l=-\infty}^{+\infty} \! S_{h, j, l} X_{h} X_{j} X_{l}^{*}, 
\end{equation}
where $\gamma$ is the nonlinear coefficient. The product of the symbol triplets are weighted by complex perturbation terms $S_{h, j, l}$ that quantify self channel interference\footnote{The cross-channel interference can be expressed in the same form as the symbol triplet product times the complex perturbation terms (see \cite[Eq.~(60)]{mecozzi2012nonlinear}).}. In \eqref{KnMd0} we use the notation $\boldsymbol{X}_{-\infty}^{+\infty}$ to emphasize that \eqref{KnMd0} is in theory an infinite-memory channel.

The perturbation terms $S_{h, j, l}$ exhibit very different magnitudes depending on their indices (see for example \cite[Fig.~5]{tao2011multiplier}). To capture the dominant terms contributing to the NLI, we consider the terms with $j=l$ as previously done in \cite[Sec.~II-B]{dar2014shaping}. Furthermore, we then truncate the indices within the channel memory $2M$. Hence, \eqref{KnMd0} is approximated as
\begin{equation}\label{KnMd}
   Z_{\text{NLI},0}(\boldsymbol{X}_{-M}^{+M})\approx \jmath \gamma \, \sum_{h=-M}^{M} X_{h} \sum_{l=-M}^{M} S_{h, l, l} |X_{l}|^2,
\end{equation}
which shows that the NLI noise $Z_{\text{NLI},0}$ depends on the weighted symbol energies within the vector of $\boldsymbol{X}_{-M}^{+M}$.

For the transmission of \emph{non}-i.i.d. symbol sequences, the sum of the ASE and the NLI in \eqref{yNLI} is approximated in \cite[Eq.~(8)]{agrell2014capacity} as
\begin{equation}\label{FEGN}
   Z_{\text{ASE},0}+Z_{\text{NLI},0}(\boldsymbol{X}_{-M}^{+M}) \approx \tilde{Z}_{0} \sqrt{P_{\mathrm{ASE}}+\eta\left(\frac{\sum_{k=-M}^{M}\left|X_{k}\right|^{2}}{2M+1} \right)^{3}},
\end{equation}
where $\tilde{Z}_0$ is a zero-mean unit-variance circularly symmetric complex Gaussian random variable, $P_{\mathrm{ASE}}$ is the ASE noise variance, and $\eta$ is a real, non-negative constant quantifying the NLI. When compared to \eqref{KnMd}, \eqref{FEGN} is a more radical approximation in the sense that the energies of all interfering symbols within a window of length $2M+1$ are assumed to make equal contribution to NLI.

In general, the expressions \eqref{KnMd} and \eqref{FEGN} show that the sum of (weighted) symbol energies within a window plays an important role in determining the NLI. Transmitting correlated symbols can therefore suppress the NLI. These observations motivate us to focus on the symbol energies and their temporal correlations.

\subsection{Related Metrics for NLI}\label{subsec:NLIMetric}
One well-known metric for showing variations of symbol energies is peak-to-average power ratio (PAPR). PAPR has been widely used for analysis of orthogonal frequency-division multiplexing (OFDM) fiber transmission \cite{armstrong2009ofdm}. PAPR partially reflects energy variation of symbols and thus can be regarded as a rough indication of nonlinearity tolerance of the transmitted signal \cite[Sec.~II-D]{chen2020analysis}, \cite[Sec.~II]{geller2016shaping}. PAPR is defined as
% PAPR
\begin{equation}
   \Theta \triangleq \frac{|X|_{\text{max}}^2}{\Exp{|X|^2}},
\end{equation}where $|X|_{\text{max}}^2$ is the maximum symbol energy in the symbol sequence.

%% kurtosis
Another important metric (originated from the EGN model) is kurtosis. The EGN model predicts strong NLI if the transmitted symbols sequence has high kurtosis. For zero-mean constellations, kurtosis is defined as
\begin{equation}\label{Kur}
  \Phi\triangleq\frac{\Exp{|X|^{4}}}{\Exp{|X|^{2}}^{2}}.
\end{equation}

% Run ratio
To explain the blocklength dependent effective SNR, run ratio was proposed in \cite[Sec.~III-B]{fehenberger2019analysis}. Symbol sequences with high run ratio are considered to be more prone to the NLI. Run ratio is defined as
\begin{equation}
\text{R}_r \triangleq\frac{1}{T}\left(1+\sum_{i=1}^{T-1} \bar{\delta}\left(x_{i-1}, x_{i}\right)\right),
\end{equation}
where $T$ is the number of transmitted symbols, and $\bar{\delta}(x_{i-1}, x_{i})$ is a decision function which is equal to $1$ when $x_{i-1} \neq x_{i}$ or $0$ otherwise.

%%%%  Sec  %%%%%%%%%%%%%%%%%%%%%%%%%%%%%%%%%%
\section{QAM Symbols Shaped by CCDM}\label{sec:CCQAM}
This section first reviews the generation of QAM symbol sequences in the PAS architecture. We then show that the symbols generated by CCDM are statistically dependent from each other. Motivated by results in Sec. II-B, which showed that the correlation among symbols could lead to a reduced NLI, we propose to analyze the autocorrelation of the symbol energies. This analysis will be later used in Sec.~IV for the derivation of the EDI.

\begin{figure}[!t]
% \resizebox{1\linewidth}{!}{\input{./Figures/PAS_system.tikz}}
\includegraphics[width=0.95\linewidth]{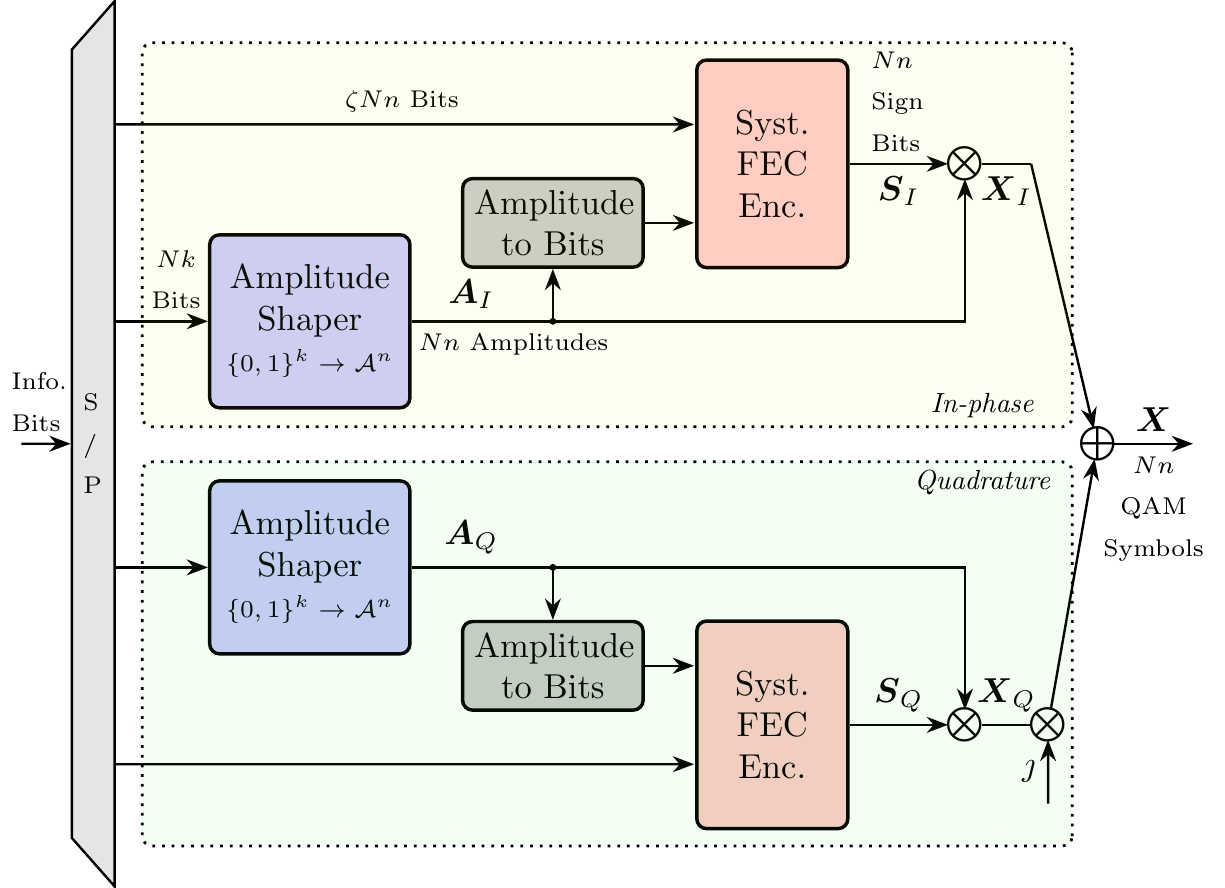}
\centering
\caption{Transmitter schematic diagram for a PAS architecture using 1D symbol mapping.}
\label{PASsys}
\end{figure}

\begin{figure*}[!t]
\centering
% \resizebox{1\linewidth}{!}{\input{./Figures/depAmpBlk.tikz}}
\includegraphics[width=0.98\linewidth]{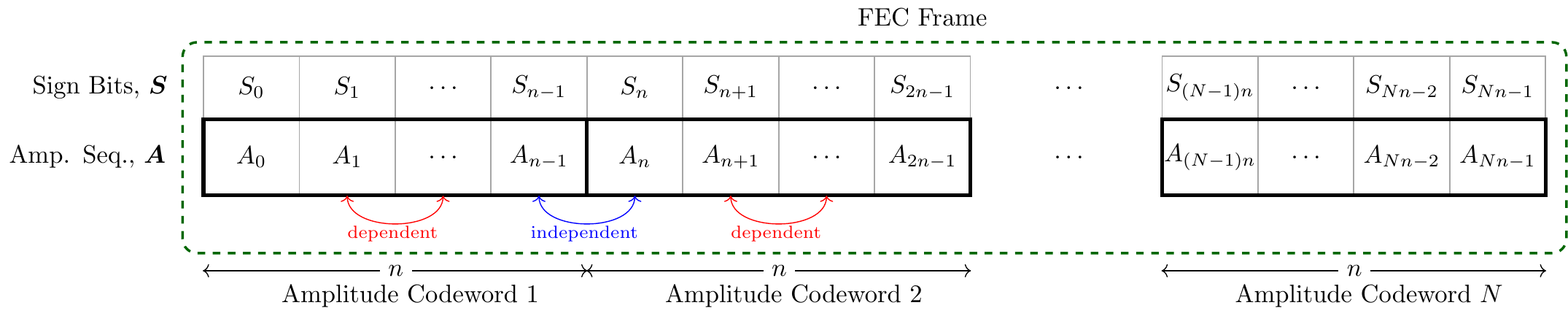}
\caption{Illustration of FEC frame and amplitude dependency. One FEC frame is formed by $nN$ amplitudes and $nN$ sign bits. Amplitudes within the same codeword are dependent, while amplitudes from different codewords are independent.}
\label{ampDep}
\end{figure*}

%%%% Subsec %%%%%%%%%%%%%%%%%%%%%%%%%%%%%%
A block diagram of a PAS transmitter is illustrated in Fig.~\ref{PASsys}. To obtain a nonuniformly distributed QAM symbol sequence $\boldsymbol{X}$, two sequences of shaped pulse-amplitude modulation (PAM) symbols $\boldsymbol{X}_{I}$ and $\boldsymbol{X}_{Q}$ are independently generated in the in-phase and quadrature dimensions, respectively. The resulting QAM symbols are then $\boldsymbol{X}=\boldsymbol{X}_{I}+\jmath\boldsymbol{X}_{Q}$. In this paper, similar to \cite[Sec.~II-C]{skvortcov2020huffman}, we refer this mapping strategy as 1D symbol mapping\footnote{This mapping is also called \emph{inter-pairing} in \cite[Sec.~II-B]{fehenberger2019analysis}.}. When obtained from independent information bit sequences, $\boldsymbol{X}_{I}$ and $\boldsymbol{X}_{Q}$ are also independent of each other.

\subsection{System Model}
As shown in Fig.~\ref{PASsys}, the generation of PAM symbols relies on a systematic FEC engine and an amplitude shaper. A fixed-length amplitude shaper encodes $k$ information bits to an \emph{amplitude codeword} of blocklength $n$, and thus the shaping rate is $k/n$. The amplitude set is denoted as $\mathcal{A}=\{\Delta,3\Delta,5\Delta,\ldots\}$ where $\Delta$ is a scaling factor.

For simplicity of exposition, let us consider the in-phase dimension and one FEC codeword. As shown in Fig.~\ref{PASsys}, a FEC codeword consists of $N$ amplitude codewords. After a serial to parallel conversion, $Nk$ information bits go into the amplitude shaper, and $\zeta Nn$ bits go into FEC, where $\zeta$ denotes the fraction of information in sign bits (see \cite[Eq.~(29)]{bocherer2015bandwidth}). The amplitude shaper converts $Nk$ bits into $N$ amplitude codewords, which we call an \emph{amplitude sequence} $\boldsymbol{A}_I=[A_{0},A_{1},\ldots,A_{Nn-1}]$. The amplitudes in $\boldsymbol{A}_I$ are then labeled into bits (see Fig.~\ref{PASsys}) and fed into the FEC along with $\zeta Nn$ bits. The labeling used is the binary-reflected Gray code \cite{frank1953pulse}. The FEC parity bits together with $\zeta Nn$ bits serve as sign bits $\boldsymbol{S}_I$ to yield the PAM symbol sequence $\boldsymbol{X}_I=[X_{I,0},X_{I,1},\ldots,X_{I,Nn-1}]$, where $X_{I,i}=(-1)^{S_{I,i}}A_{I,i}$ ($i=0,1,\ldots,Nn-1$). The same procedure is conducted in the quadrature branch to generate $\boldsymbol{X}_Q$. 

The amplitude shaper imposes constraints on the amplitudes of each amplitude codeword. In this paper, the amplitude shaper we consider is CCDM. Due to the constant-composition (CC) constraint, given blocklength $n$, the number of amplitudes $a\in\mathcal{A}$ in every codeword is a constant denoted by $n_{a}$, and thus $n = \sum_{a\in\mathcal{A}} n_{a}$. For any realization of the amplitude sequence, the amplitude frequency distribution of $a$ is $n_{a}/n$. Therefore, the amplitude codewords generated by CCDM are simply permuted versions of each other.

In this paper, we make two assumptions about the CCDM codebook. In the following section, these assumptions will allow us to consider the CCDM codewords as a genuine process of drawing amplitudes without replacement \cite[Sec.~IV]{schulte2015constant}. The first assumption is that the codewords are equally likely, which is justified in the scenario where the information bits are independent and uniformly distributed. The second assumption is that the CCDM codebook contains all possible amplitude permutations. The total number of permutations is given by the multinomial coefficient of the composition (see \cite[Eq.~(5)]{gultekin2019probabilistic}), which is denoted as $N_C$. For $k$ input bits, CCDM selects $2^k$ permutations as codewords. In general, $2^k < N_C$ because (i) $k$ is chosen to be variable to for example achieve rate adaptivity, or (ii) $N_C$ usually not be exactly a power of two. In this paper, we always use the largest possible value of $k$, and thus, (i) above can be ignored. On the other hand, due to (ii) above, CCDM does not always generate all $N_C$ permutations. In this paper, we use \emph{emulated} CCDM codewords that include all amplitude permutations. We will show later in the paper that this approximation is very good when compared to ``exact'' CCDM, where not all permutations are used.

%\ns{In this paper, we mainly use emulated CCDM codewords that include all amplitude permutations. By contrast, to yield an integer value of $k$, genuine CCDM codewords are affected by the rounding of $\log_2(N_C)$, thereby losing a number of amplitude permutations. We will also show that using genuine CCDM codewords with the maximum shaping rate has negligible effect on our analytical results, even if not all permutations are used.}
%This assumption can be justified by the fact that PAS usually keep the maximum shaping rate that approaches the entropy of amplitude distribution. achieves rate granularity normally through adjusting the amplitude composition, while keeping the maximum shaping rate $k/n$. This means that for sufficiently long blocklength $n$, almost all permutations will be used as codewords (see \cite[Eq.~(37)]{bocherer2015bandwidth}). 

\subsection{Statistical Dependency Among CC Amplitudes}
The dependency of amplitudes in a FEC frame is illustrated in Fig.~\ref{ampDep}. For notation simplicity, in what follows we drop the subscripts of $\boldsymbol{A}$ and $\boldsymbol{S}$ that indicate in-phase or quadrature. Furthermore, we pay little attention to sign bits since they only determine the polarity of PAM symbols. In terms of the amplitude sequence $\boldsymbol{A}$, Fig.~\ref{ampDep} schematically shows that the amplitudes from different amplitude codewords of length $n$ are independent of each other. Fig.~\ref{ampDep} also shows that amplitudes \emph{within} an amplitude codeword are mutually \emph{dependent}.  

The amplitude dependency within amplitude codewords comes from the CC constraint. During the arithmetic encoding of a CCDM that can be modeled as drawing without replacement process, the probability of drawing an amplitude is always updated by subtracting the previous number of amplitudes from the composition. The CC constraint further implicitly ensures that the sum of amplitudes in a CCDM codeword is a constant, i.e.,
\begin{equation}\label{linear}
    \sum_{j=0}^{n-1}A_j = \sum_{a\in\mathcal{A}} an_{a}.
\end{equation}
For any two different time instants $i$ and $i+\tau$, we then have
\begin{equation}\label{linear2}
A_i  =-A_{i+\tau}-\sum_{\substack{j=0\\ j\neq i, i+\tau}}^{n-1}A_j + \sum_{a\in\mathcal{A}} a n_{a}.
\end{equation}
Expression \eqref{linear2} shows that these amplitudes $A_i$ and $A_{i+\tau}$ within CCDM codewords exhibit a \emph{linear relationship}. This observation shows that amplitudes within an amplitude codeword are indeed linearly dependent.

In the next example, we use a CCDM shaping trellis to explain the statistical dependency between amplitudes. The trellis shows all possible amplitude sequences and their accumulated energies. We construct the trellis based on the following rules:
\begin{itemize}
    \item Each path in the trellis represents a unique amplitude sequence.
    \item At time instant $i$, the vertical position of the nodes represents the accumulated energy $\AcEnergy_{i}\in \mathcal{E}$, where
    \begin{equation}
        \AcEnergy_{i} \triangleq \sum_{t=0}^{i-1}|A_t|^2, 
    \end{equation}
    $\AcEnergy_{0}=0$, and $\mathcal{E}$ is the set of possible accumulated energy levels.
    \item The numbers labeling the trellis states represent the value of $\Prob_{A_{i}\mid \AcEnergy_{i}}(a|e)$, i.e., the conditional probability of amplitude $A_{i}=a$ given the accumulated energy.%\ns{\footnote{Note that an accumulated energy node can be reached by paths with different compositions, which is not considered in Example~\ref{Example1}.}}%Note that for the process $A_{i}$, unlike Markov chains, the conditional probability $\Prob_{A_{i}\mid E_{i}}(a|e)$ in general also depends on the specific path reaching the node at trellis section $i$ with energy state $E_i$. However, this dependency does not appear in the specific case illustrated in Example~\ref{Example1}.} 
    $\AcEnergy_i=e$ for $i=0,1,\ldots$, where $a\in\mathcal{A}$ and $e\in\mathcal{E}$. 
    \item At time instant $i$, the edges indicate the amplitude $A_i$. The steeper the slope of the edge has, the larger the amplitude. The number next to the edges is $\Prob_{A_{i},\AcEnergy_{i}}(a, e)$, i.e., the probability of the paths starting from accumulated energy $e$ using an amplitude $a$.
\end{itemize}

\begin{figure}[!t]
    \centering
    % \resizebox{0.8\linewidth}{!}{\input{./Figures/depExample.tikz}}
    \includegraphics[width=0.65\linewidth]{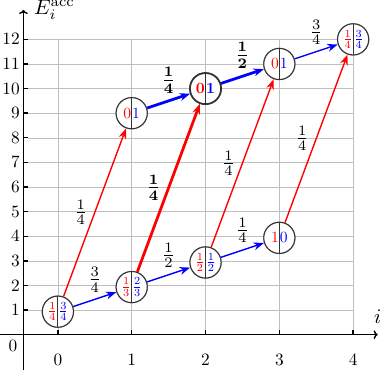}
    \caption{CCDM shaping trellis with blocklength $n=4$ for amplitude set $\mathcal{A}=\{1,3\}$ with $\Prob_{A}=[\frac{3}{4},\frac{1}{4}]$. The amplitude edges are labeled with $\Prob_{A_{i},\AcEnergy_{i}}(a,e)$. Each node contains $\Prob_{A_{i}\mid \AcEnergy_{i}}(3|e)$ on the left (in red) and $\Prob_{A_{i}\mid \AcEnergy_{i}}(1|e)$ on the right (in blue).}
\label{depExample1}
\end{figure}

\begin{example}[CCDM Trellis]\label{Example1}
Assume the use of CCDM with blocklength $n=4$ and a composition of three amplitudes 1 and one amplitude 3 (i.e., $n_1=3$ and $n_3=1$), and thus, the amplitude PMF is $\Prob_{A}=[\frac{3}{4},\frac{1}{4}]$. Fig.~\ref{depExample1} shows the trellis of the first amplitude codeword, whose behavior will repeat for the subsequent codewords. The set of accumulated energy states is $\mathcal{E}=\{0,1,2,3,9,10,11,12\}$. To explain the trellis, consider state $\AcEnergy_2=10$ and the three connected edges (shown with boldface text in Fig.~\ref{depExample1}). Since two paths reach this state, $\Prob_{\AcEnergy_{2}}(10)=\Prob_{A_1, \AcEnergy_1}(3,1)+\Prob_{A_1, \AcEnergy_1}(1,9)=\frac{1}{4}+\frac{1}{4}=\frac{1}{2}$. At this node, we can only choose amplitude 1 for $a_2$, and thus $\Prob_{A_{2}\mid \AcEnergy_{2}}(3|10)=0$ and $\Prob_{A_{2}\mid \AcEnergy_{2}}(1| 10)=1$. The joint probability of amplitude $a_2=1$ starting from $e_2=10$ is thus $\Prob_{A_2,\AcEnergy_{2}}(1,10)=\Prob_{A_2\mid \AcEnergy_{2}}(1| 10)\:\Prob_{\AcEnergy_{2}}(10)=1\times\frac{1}{2}=\frac{1}{2}$. 
\end{example}

In Example~\ref{Example1}, the amplitudes within a CCDM codeword are statistically dependent among each other. This can be seen from Fig.~\ref{depExample1}, where $\Prob_{A_{i}\mid \AcEnergy_{i}}(a|e)= \Prob_{A_{i}}(a)$ is not always satisfied. The reason for this dependency is that $A_0+A_1+A_2+A_3=6$ (see \eqref{linear}), and any pair of amplitudes exhibits a linear relationship. On the other hand, for i.i.d. amplitude sequences, $\Prob_{A_{i}\mid \AcEnergy_{i}}(a| e)=\Prob_{A_i}(a)$ is satisfied for $\forall a \in \mathcal{A}$.

Another property of i.i.d. amplitude sequences is the time-independent marginal probability of an amplitude. As can be seen in Example~\ref{Example1}, the marginal probability of an amplitude within a CCDM codeword is also constant, i.e., for $\forall i \in \mathbb{Z}$,
\begin{align}\label{A.stationary}
    \sum_{e\in\mathcal{E}}\Prob_{A_{i},\AcEnergy_{i}}(a, e)=\Prob_{A_i}(a)=\Prob_{A}(a)=\frac{n_a}{n}.
\end{align}
The property in \eqref{A.stationary} is due to the set of codewords being permutation invariant, as shown in \cite[Lemma~1]{yunus2020comparison}.
For example, $\Prob_{A_2}(1)=\Prob_{A_2,\AcEnergy_{2}}(1,10)+\Prob_{A_2,\AcEnergy_{2}}(1,2)=\frac{1}{2}+\frac{1}{4}=\frac{3}{4}$. The property in \eqref{A.stationary} holds for all CC amplitude sequences and will be be used in the statistical analysis later in the paper.

A closer look at Example~\ref{Example1} also reveals that when multiple CCDM codewords are transmitted, the trellis extends in the time domain, and the probabilistic model of the first amplitude codeword repeats with period $n$. This repetition is studied in the next example.

\begin{figure}[t!]
\centering
\setkeys{Gin}{width=0.24\textwidth}
\subfloat[Trellis for $n=4$.
\label{depExample_b}]{
% \resizebox{0.48\linewidth}{!}{\input{./Figures/depExample2.tikz}}
\includegraphics[width=0.46\linewidth]{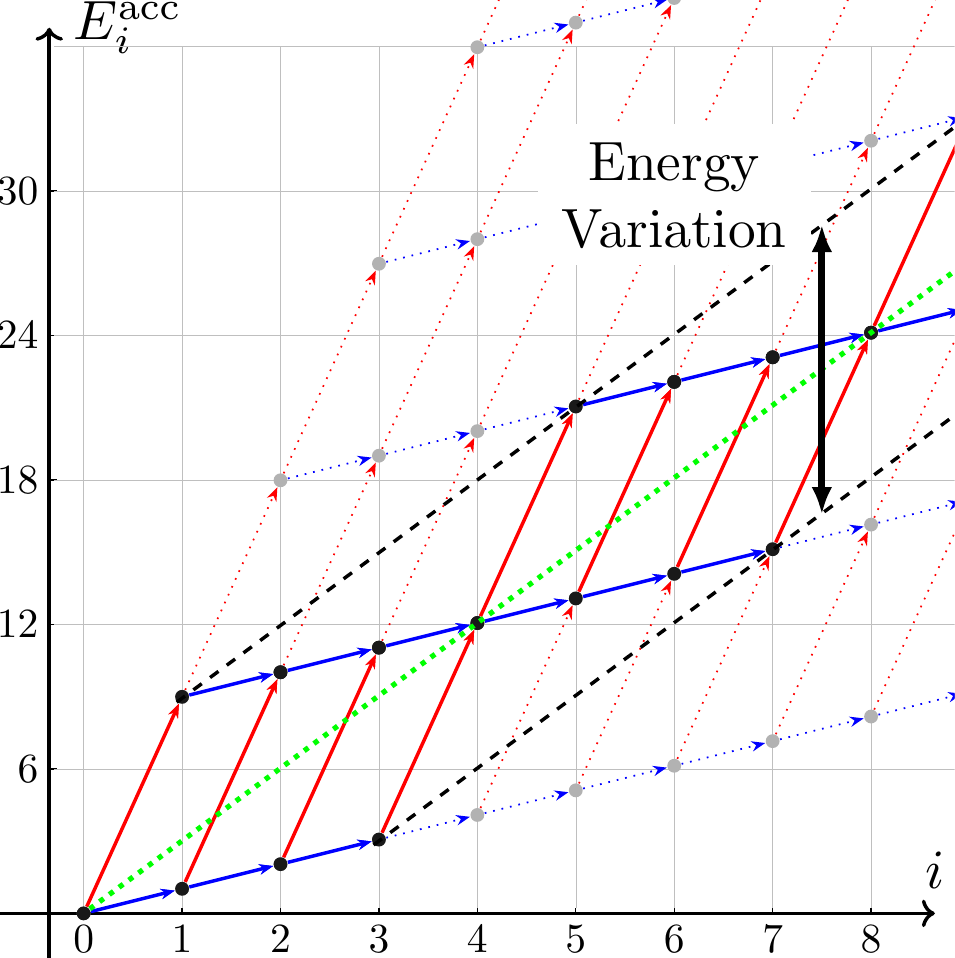}
}
\hfill
\subfloat[Trellis for $n=8$.
\label{depExample_c}]{
% \resizebox{0.48\linewidth}{!}{\input{./Figures/depExample3.tikz}}
\includegraphics[width=0.46\linewidth]{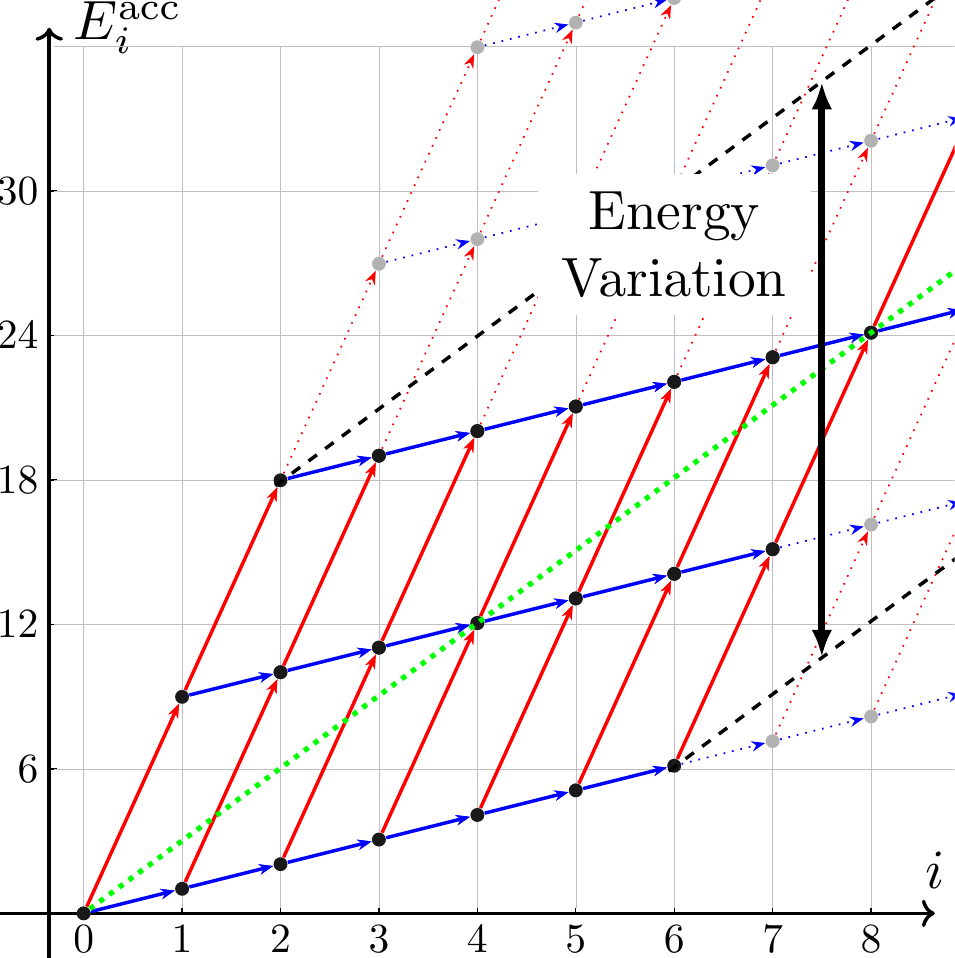}
}
\caption{CCDM shaping trellis comparison for blocklengths $n=4$ and $n=8$, with $\Prob_{A}=[\frac{3}{4},\frac{1}{4}]$. The trellis marked by gray dotted lines show amplitude sequences never generated by CCDM.}
\label{depExample2}
\end{figure}

%\textit{Example 2:} 
\begin{example}[Trellis Repetition and Blocklength]\label{Example2}
Fig.~\ref{depExample2} shows a comparison of the CCDM shaping trellises for blocklengths $n=4$ and $n=8$. Since both cases have the same $\Prob_{A}$, it can be seen that $\AcEnergy_4/4=12/4=3$ for $n=4$, which is equal to $\AcEnergy_8/8=24/8=3$ for $n=8$ (see Fig.~\ref{depExample2}). As a result, the green dotted lines in the middle of the trellises show that their accumulated energies grow at the same average rate. We also note that for blocklength $n=8$, the trellis broadens (in the energy direction) and includes the trellis for $n=4$. The vertical arrows between the black dashed lines shows that the variations of the accumulated energy growth for $n=8$ is larger when compared to $n=4$.
\end{example}

An important implication from the two examples above is that by introducing dependencies between amplitudes, CCDM artificially constrains the \emph{dispersion} of the accumulated energy of the amplitude sequences (see black arrows in Fig.~\ref{depExample2}). For a fixed PMF $\Prob_{A}$, the accumulated energy dispersion is more constrained with shorter blocklength.

\subsection{Statistics of Symbol Energies with CC Amplitudes}

In this section, we will study the statistical properties of the QAM symbol sequences defined by CC amplitudes from the perspective of stochastic processes. In particular, we will show that the process of \emph{symbol energies} is wide-sense cyclostationary (WSCS). The analysis in this section will lay the ground for the definition and formulation of the EDI metric in Sec.~\ref{sec:EDI}. We start by briefly reviewing some of the key properties of stochastic processes.

\begin{definition}[{First-order stationarity \cite[Ch.~9-1]{papoulis2002probability}}]\label{DefFos}
A stochastic process $\boldsymbol{X}$ is called \emph{first-order stationary} if $\Prob_{X_i}=\Prob_{X_{i+\tau}}$ holds for any delay $\tau$.
\end{definition}

\begin{definition}[{Autocorrelation \cite[Def.~13.13]{yates2014probability}}]\label{DefAuto}
The autocorrelation of a stochastic process $\boldsymbol{X}$ as a function of time slot $i$ and delay $\tau$ is defined as
\begin{equation}
    R_{X}(i,\tau) \triangleq \Exp{X_i X_{i+\tau}}. \label{AutoCorrDef0}
\end{equation}
\end{definition}

\begin{definition}[{Autocovariance \cite[Def.~13.12]{yates2014probability}}]\label{DefAutoCov}
The autocovariance of a stochastic process $\boldsymbol{X}$ is defined as
\begin{equation}
  \Cov{X_i,X_{i+\tau}} \triangleq  \Exp{\left(X_i-\Exp{X_i}\right)\left(X_{i+\tau}-\Exp{X_{i+\tau}}\right)}.\label{EqDefAutoCov} 
\end{equation}
Moreover, for first-order stationary processes,
\begin{equation}
  \Cov{X_i,X_{i+\tau}} = R_{X}(i,\tau)-\Exp{X}^2.\label{DefAutoCovStationary} 
\end{equation}
\end{definition}

\begin{definition}[{Wide-sense Cyclostationarity \cite[Ch.~10-4]{papoulis2002probability}}]\label{DefWSCS}
A stochastic process $\boldsymbol{X}$ is called \emph{wide-sense cyclostationary} with period $n$ if $\Exp{X_i}=\Exp{X_{i+mn}}$ and $R_{X}(i,\tau)=R_{X}(i+mn,\tau)$ hold for every $m\in\mathbb{Z}$.
\end{definition}

In the definitions above, we used $\boldsymbol{X}$ to denote a generic stochastic process. In what follows, we focus on the process of symbol energies defined as
\begin{equation}\label{EnergySeqDef}
\Energies \triangleq [\ldots,\Energy_{i-1},\Energy_i,\Energy_{i+1},\ldots],
\end{equation}
where
\begin{equation}\label{EnergyDef}
\Energy_i \triangleq |X_i|^2=A_{I,i}^2+A_{Q,i}^2.
\end{equation}
For a symbol energy process $\Energies$, we will refer to every $n$ symbol energies as a \emph{block}\footnote{We avoid using ``symbol energy codeword'', since the mapping between $k$ input bits and $n$ symbol energies is not one-to-one.}, which are constructed based on two amplitude codewords (see Fig.~\ref{PASsys}). In addition, the autocorrelation of $\Energies$ is
\begin{equation}\label{autoCorrEnergy}
    R_{\Energy}(i,\tau) \triangleq \Exp{\Energy_i\Energy_{i+\tau}}.
\end{equation}

\begin{lemma}\label{Thmmin1}
The QAM symbol energies $\Energies$ defined by CC amplitudes is a first-order stationary process.
\end{lemma}
\begin{IEEEproof}
As shown in \eqref{A.stationary}, the CC amplitude sequences $\boldsymbol{A}$ are first-order stationary processes. Since a QAM symbol can be decomposed into two independent dimensions, the process of the energies $\Energies$ in \eqref{EnergySeqDef} is also first-order stationary.
\end{IEEEproof}

Lemma~\ref{Thmmin1} shows that the probability distribution of symbol energy $\Prob_{\Energy_i}$ is constant for any $i\in\mathbb{Z}$. From this first-order stationarity property in Lemma~\ref{Thmmin1}, it follows that
\begin{align}
   \Exp{\Energy_i} =\; & \Exp{|X_i|^2} = \Exp{|X|^2}, \label{EX2Stat} \\
   \Var{\Energy_i} =\; &\Var{|X_i|^2} = \Var{|X|^2},\label{VarStat}
\end{align}
where
\begin{align}
 \Exp{|X|^2} =\; & 2\Exp{A^2}, \label{EX2}\\
 \Var{|X|^2} =\; & 2\Exp{A^4} -  2 \Exp{A^2}^2, \label{VX2}
\end{align}
and \eqref{VX2} was obtained using
\begin{equation}
    \Exp{|X|^4} = 2\Exp{A^4} + 2 \Exp{A^2}^2. \label{EX4}
\end{equation}
The second and fourth order moments of $A$ in \eqref{EX2}--\eqref{EX4} can be computed by using the amplitude PMF $\Prob_{A}(a)=n_a/n$, i.e.,
\begin{align}
\Exp{A^2} =\sum_{a\in\mathcal{A}} a^2 \frac{n_{a}}{n}, \label{EA2}\\
 \Exp{A^4} =\sum_{a\in\mathcal{A}} a^4\frac{n_{a}}{n}. \label{EA4}
\end{align}

\begin{theorem}\label{Thm:NegCov}
For the sequence of QAM symbol energies $\Energies$ defined by CC amplitudes, if $\Energy_{i}$ and $\Energy_{i+\tau}$ ($\tau\neq 0$) belong to the same block, we have
\begin{align}
    R_{\Energy}(i,\tau)  = \CrS  & < \Exp{|X|^2}^2, \label{bitcorr1}
\end{align}
where
\begin{align}
   \CrS  &  \triangleq \frac{n\Exp{|X|^2}^2-\Exp{|X|^4}}{n-1}, \label{bitcorr0}
\end{align}
and
\begin{align}
    \Cov{E_i,E_{i+\tau}} & = -\frac{\Var{|X|^2}}{n-1}  < 0 . \label{negCov}
\end{align}

\end{theorem}
\begin{IEEEproof}
See Appendix~\ref{Proof:NegCov}.
\end{IEEEproof}

In Theorem~\ref{Thm:NegCov}, the negativity of autocovariance in \eqref{negCov} shows that the dependency between amplitudes is due to the fact that these amplitudes are inversely correlated. This negativity will also be used to prove Corollary~\ref{Thm:LowLmt} in Sec.~\ref{sec:EDI}. In addition, from \eqref{negCov} we obtain $\lim_{n\to\infty}\Cov{E_i,E_{i+\tau}}=0$, indicating that using long blocklengths weakens the linear dependency between symbol energies. %Furthermore, we emphasize that $\CrS$ in \eqref{bitcorr1} corresponds to $R_{\Energy}(i,\tau)$ for the case when $\Energy_{i}$ and $\Energy_{i+\tau}$ belong to the same block.

\begin{lemma}\label{Thm0}
The QAM symbol energies $\Energies$ defined by CC amplitudes is a WSCS process.
\end{lemma}
\begin{IEEEproof}
The first condition of WSCS processes $\Exp{\Energy_i}=\Exp{\Energy_{i+mn}}$ is satisfied because of \eqref{EX2Stat}. The second condition of WSCS processes $R_{\Energy}(i,\tau)=R_{\Energy}(i+mn,\tau)$ follows from the fact the same probabilistic model repeats for every block of $n$ symbols (see Fig.~\ref{ampDep}).
\end{IEEEproof} 

In practice, we are interested in the average autocorrelation of WSCS processes over the cyclostationarity period. For the process under investigation, such an average autocorrelation $\AC$ is defined as
\begin{align}
    \AC & \triangleq \frac{1}{n} \sum_{i=0}^{n-1} R_{\Energy}(i,\tau) \label{AutoCorrDef} \\
    & = \frac{1}{n} \sum_{i=0}^{n-1} \Exp{\Energy_i \Energy_{i+\tau}}, \label{AutoCorrDef2}
\end{align} 
where \eqref{AutoCorrDef2} follows from \eqref{autoCorrEnergy}. The average autocorrelation $\AC$ in \eqref{AutoCorrDef} can be interpreted as a quantity that indicates the average linear dependency of all possible pairs of symbol energies separated by $\tau$ symbols.
 
The following theorem gives an analytical expression for the average autocorrelation in \eqref{AutoCorrDef2} for the considered QAM symbol sequences.

% correlation table
%%% Table correlation contribution
\begin{table*}[!t]
    \centering
    \caption{The autocorrelation $R_{\Energy}(i,\tau)$ and its average $\AC$ for $\tau>0$. The three cases of $R_{\Energy}(i,\tau)$ in \eqref{AutoCorrSplt} correspond to the first column (in gray), the upper left part (in blue) and the bottom right (in red), respectively.}
    \label{CorrContb}
    \begin{tabular}{l|ccccccccc}
    %\begin{tabular}{m{0.8cm}|m{1.2cm}m{1.8cm}m{1.8cm}m{1cm}m{1.8cm}m{1.8cm}m{1.2cm}m{1.2cm}m{1.2cm}m{1.2cm}}
\hline 
 & \multicolumn{9}{c}{$\tau$} \\
\cline{2-10} 
$i$ & $0$ & $1$ & $2$ & $\ldots$ & $n-2$ & $n-1$ & $n$ & $n+1$ & $\ldots$ \\
\hline 
$0$ & \cellcolor{lightgray} $\Exp{|X|^4}$ & \cellcolor{blue!20}  $\CrS$ & \cellcolor{blue!20}  $\CrS$ & \cellcolor{blue!20}$\ldots$ & \cellcolor{blue!20}  $\CrS$ & \cellcolor{blue!20}  $\CrS$ & \cellcolor{red!20}  $\Exp{|X|^2}^2$  & \cellcolor{red!20}  $\Exp{|X|^2}^2$ & \cellcolor{red!20}$\ldots$\\
$1$ & \cellcolor{lightgray} $\Exp{|X|^4}$ & \cellcolor{blue!20}  $\CrS$ & \cellcolor{blue!20}  $\CrS$ & \cellcolor{blue!20}$\ldots$ & \cellcolor{blue!20}  $\CrS$ & \cellcolor{red!20}  $\Exp{|X|^2}^2$ & \cellcolor{red!20}  $\Exp{|X|^2}^2$ & \cellcolor{red!20}  $\Exp{|X|^2}^2$ & \cellcolor{red!20}$\ldots$  \\
$\vdots$ & \cellcolor{lightgray} $\vdots$ & \cellcolor{blue!20} $\vdots$ & \cellcolor{blue!20}$\vdots$ & $\ddots$ & \cellcolor{red!20}$\vdots$ & \cellcolor{red!20}$\vdots$ & \cellcolor{red!20}$\vdots$ & \cellcolor{red!20}$\vdots$& \cellcolor{red!20}$\vdots$ \\
$n-2$ & \cellcolor{lightgray} $\Exp{|X|^4}$ & \cellcolor{blue!20}  $\CrS$ & \cellcolor{red!20}  $\Exp{|X|^2}^2$ & \cellcolor{red!20}$\ldots$ & \cellcolor{red!20}  $\Exp{|X|^2}^2$ & \cellcolor{red!20}  $\Exp{|X|^2}^2$ & \cellcolor{red!20}  $\Exp{|X|^2}^2$ & \cellcolor{red!20}  $\Exp{|X|^2}^2$& \cellcolor{red!20}$\ldots$ \\
$n-1$ & \cellcolor{lightgray} $\Exp{|X|^4}$ & \cellcolor{red!20}  $\Exp{|X|^2}^2$ & \cellcolor{red!20}  $\Exp{|X|^2}^2$ & \cellcolor{red!20}$\ldots$ & \cellcolor{red!20}  $\Exp{|X|^2}^2$ & \cellcolor{red!20}  $\Exp{|X|^2}^2$ & \cellcolor{red!20}  $\Exp{|X|^2}^2$ & \cellcolor{red!20}  $\Exp{|X|^2}^2$  & \cellcolor{red!20}$\ldots$\\
\hline
\multirow{2}*{$\AC$} & \cellcolor{lightgray}  & \cellcolor{blue!20} $(n\!-\!1)\CrS/n+$ & \cellcolor{blue!20} $(n\!-\!2)\CrS/n+$ &\cellcolor{blue!20}$\ldots$ & \cellcolor{blue!20}$2\CrS/n+$ & \cellcolor{blue!20}$\CrS/n+$ & \cellcolor{red!20} & \cellcolor{red!20} & \cellcolor{red!20}$\ldots$ \\
& \multirow{-2}*{\cellcolor{lightgray}$\Exp{|X|^4}$} & \cellcolor{red!20} $\Exp{|X|^2}^2$/n & \cellcolor{red!20} $2\Exp{|X|^2}^2/n$ & \cellcolor{red!20}$\ldots$ &\cellcolor{red!20} $(n\!-\!2)\Exp{|X|^2}^2/n$ & \cellcolor{red!20} $(n\!-\!1)\Exp{|X|^2}^2/n$ & \multirow{-2}*{\cellcolor{red!20}$\Exp{|X|^2}^2$} & \multirow{-2}*{\cellcolor{red!20}$\Exp{|X|^2}^2$}  & \cellcolor{red!20}$\ldots$\\
\hline
\end{tabular}
\end{table*}

\begin{theorem}\label{Thm:AvgAutoCorrEng}
The average autocorrelation $\AC$,  $\tau\in\mathbb{Z}$ for CCDM QAM symbol sequences with blocklength $n$ generated using a composition such that $\Prob_{A}(a)=n_a/n,a\in\mathcal{A}$ can be expressed as
\begin{equation}\label{AutoCorr}
\setlength{\nulldelimiterspace}{0pt}
\AC=\left\{\begin{IEEEeqnarraybox}[\relax][c]{l's}
\Exp{|X|^4},&if $\tau=0$\\
\frac{|\tau|\Exp{|X|^2}^2+(n-|\tau|)\CrS}{n},&if $1\leq|\tau|< n$\\
\Exp{|X|^2}^2, &if $|\tau| \geq n$
\end{IEEEeqnarraybox}\right.
\end{equation}
where $\CrS$ is given in \eqref{bitcorr0}.
\end{theorem}
\begin{IEEEproof}
The function $R_{\Energy}(i,\tau)$ for different values of $\tau$ and $i$ is illustrated in Table~\ref{CorrContb}, where for simplicity only $\tau>0$ is shown. When $\tau=0$, $R_{\Energy}(i,0)=\Exp{|X|^4}$ as shown in the first column in Table~\ref{CorrContb}. For $\tau\neq0$, $\Energy_{i+\tau}$ and $\Energy_{i}$ are either at the same block and thus correlated, or at two different blocks and thus independent of each other. For example, when $i=0$, $\Energy_{0}$ is only correlated with the following $n-1$ symbol energies, yielding $R_{\Energy}(0,\tau) = \CrS$ for $\tau=1,2,\ldots,n-1$ (see Fig.~\ref{ampDep} and the blue cells in the $i=0$ row of Table~\ref{CorrContb}). When $\tau=n$, $\Energy_{n}$ is the energy of the first symbol in the second block, and thus $\Energy_{0}$ and $\Energy_{n}$ are independent that gives $R_{\Energy}(0,n) = \Exp{|X|^2}^2$. This is shown by the red cells in the $i=0$ row of Table~\ref{CorrContb}. The other rows in Table~\ref{CorrContb} can be calculated in a similar way. $\AC$ is the column-wise average of $R_{\Energy}(i,\tau)$.
\end{IEEEproof}

Note that $\AC$ is an even function and that $\AC$ in \eqref{AutoCorr} is completely determined by the 
% composition, amplitude set $\mathcal{A}$, 
blocklength $n$, the second, and fourth order moments of $|X|$. The latter can be calculated via \eqref{EX2}--\eqref{VX2} and \eqref{EA2}--\eqref{EA4}.

Furthermore, $\AC$ can be approximated by the sample autocorrelation function $\TAC$ in a Monte Carlo simulation. With $T$ samples, $\TAC$ is defined as
\begin{equation}
    \TAC \triangleq \frac{1}{T} \sum_{t=0}^{T-1}|x_t|^2|x_{t+\tau}|^2. \label{AutoCorrEst}
\end{equation}
When $T\to\infty$, $\TAC\to\AC$. This ergodicity follows from Slutsky's theorem \cite[Thm.~12-2]{papoulis2002probability}, due to the fact that as $\tau'\to\infty$, the correlation between symbol energies vanishes and thus $\Cov{\Energy_i^2\Energy_{i+\tau}^2,\Energy_{i+\tau'}^2\Energy_{i+\tau+\tau'}^2}\to 0$.
%In the following example, we will use $\AC$ as an indicator of statistical dependence between the symbol energies at different time slots. 

\begin{figure}[!t]\label{Corr}
    \centering
    % \resizebox{1\linewidth}{!}{\input{./Figures/foMOrdErr_COMP.tikz}}
    \includegraphics[width=1\linewidth]{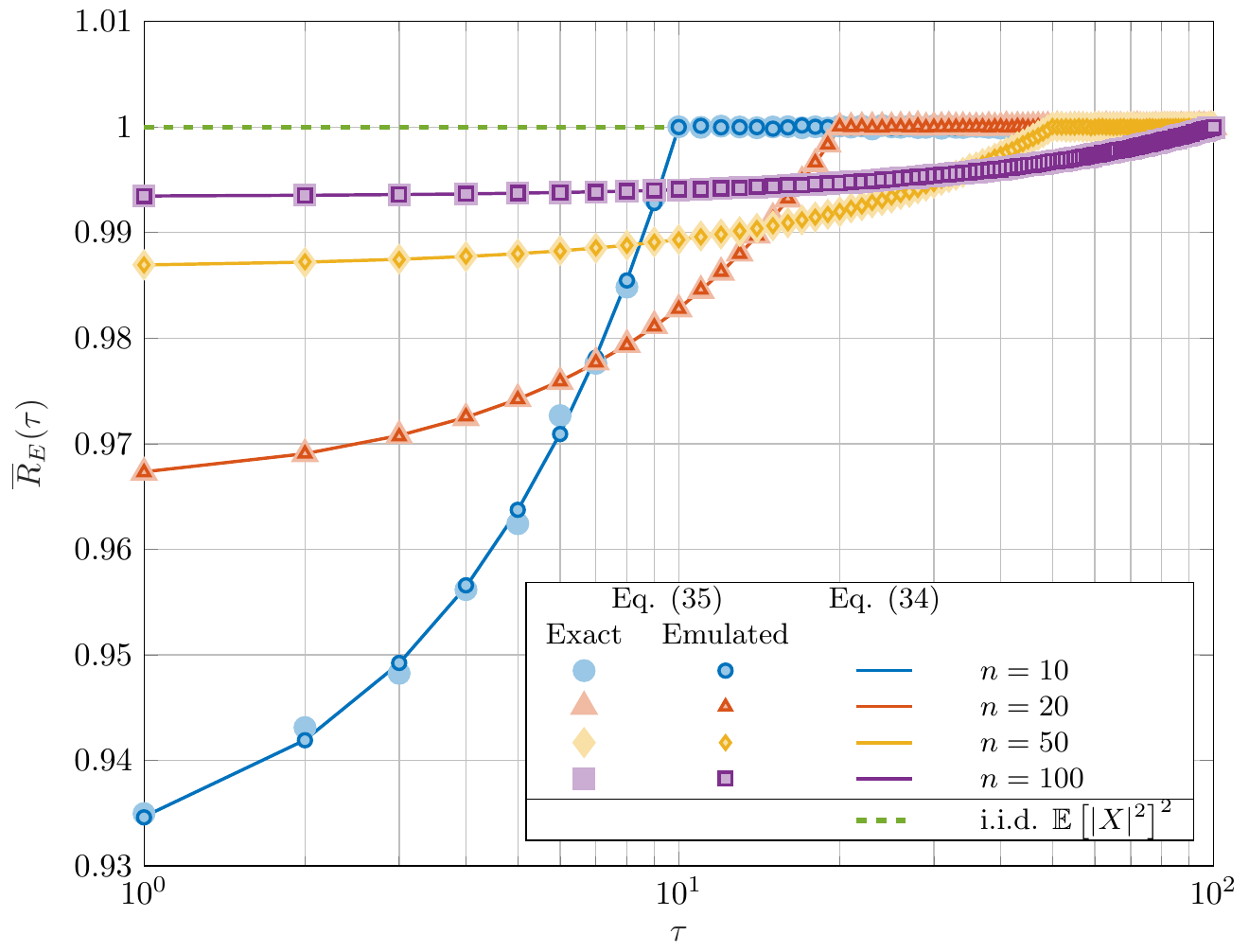}
    \caption{Simulation results of $\TAC$ in \eqref{AutoCorrEst} (markers) by using exact or emulated CCDM codewords, and analytical results of $\AC$ given by Theorem~\ref{Thm:AvgAutoCorrEng} (solid lines) for normalized $\left(\Exp{|X|^2}=1\right)$ 64QAM symbol sequences with $\Prob_{A}=[0.4,0.3,0.2,0.1]$. For all cases $\overline{R}_{E}(0)=\Exp{|X|^4}=1.65$.}
    \label{fOMErr}
\end{figure}
%\textit{Example 3:} 
\begin{example}[Average Autocorrelation]\label{Example3}
Fig.~\ref{fOMErr} shows the average autocorrelation $\AC$ in Theorem~\ref{Thm:AvgAutoCorrEng} and its estimation $\TAC$ for 64QAM symbol sequences (normalized to $\Exp{|X|^2}=1$). Four different shaping blocklengths are compared, which use the same distribution $\Prob_{A}=[0.4,0.3,0.2,0.1]$, and have the same $\overline{R}_{\Energy}(0)=\Exp{|X|^4}=1.65$ (not shown in Fig.~\ref{fOMErr}). I.i.d. sequences with the same distribution are also considered. Fig.~\ref{fOMErr} shows $\TAC$ in \eqref{AutoCorrEst} for sequences generated using exact CCDM, emulated CCDM, and for a system with a symbol-level interleaver to emulate the i.i.d. property. Fig.~\ref{fOMErr} shows that the emulated results differ from exact CCDM only for $n=10$, where only $65\%$ of the total number of permutations $N_C$ are used as codewords. Except this minor mismatch for $n=10$, $\AC$ in Theorem~\ref{Thm:AvgAutoCorrEng} approximates well the true autocorrelation function $\AC$ in all other cases. It can also be seen in Fig.~\ref{fOMErr} that for i.i.d. symbol sequences or $\tau\geq n$, $\AC=\Exp{|X|^2}^2=1$. In the cases of CC symbol sequences with $0<\tau < n$, due to $R_{\Energy}(i,\tau)$ being smaller than $\Exp{|X|^2}^2$, the average symbol energy dependency manifest itself as a deviation between $\AC$ and $\Exp{|X|^2}^2$ (see \eqref{bitcorr1} and the second case in \eqref{AutoCorr}). %This can be understood based on two observations: (i) non-zero values of $\Cov(|X_i|^2,|X_{i+\tau}|^2)$ is tantamount of statistical dependence in the process $\boldsymbol{X}$; (ii) the relationship between $\Cov(|X_i|^2,|X_{i+\tau}|^2)$ and $\AC$ is given by \eqref{Cov}. 
For each blocklength, as $\tau$ increases, the symbol energy dependency gradually decreases and vanishes when $\tau=n$ such that $\Energy_i$ and $\Energy_{i+\tau}$ always belong to two independent blocks. Moreover, as $n$ increases, the curves approach $\Exp{|X|^2}^2$ (i.e., the deviation decreases), which implies a weaker dependency.
\end{example}

%%%%  Sec  %%%%%%%%%%%%%%%%%%%%%%%%%%%%%%%%%%
\section{Energy Dispersion Index}\label{sec:EDI}

\begin{figure}[!t]
    \centering
    % \resizebox{1\linewidth}{!}{\input{./Figures/WndEngExample.tikz}}
    \includegraphics[width=1\linewidth]{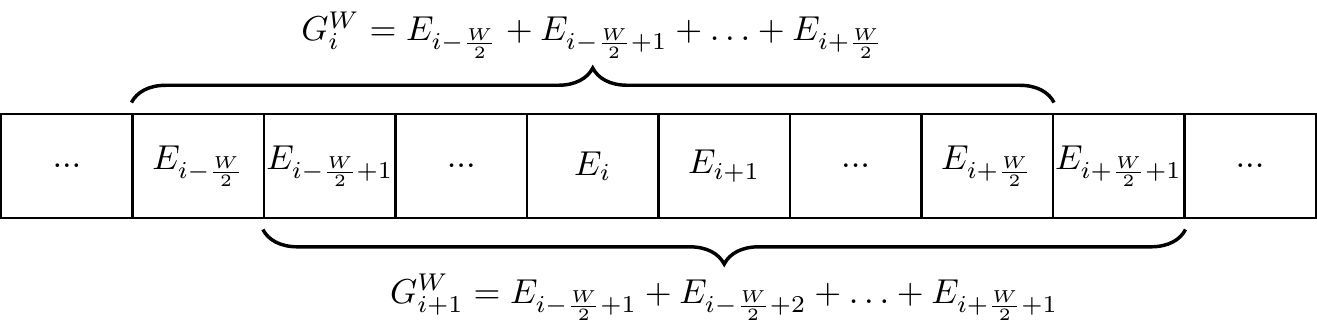}
    \caption{Window energies $\WE_i$ and $\WE_{i+1}$ given symbol energy sequence $\Energies$.}
    \label{wndEngExmp}
\end{figure}

In this section, EDI is introduced as a figure of merit to qualitatively predict the NLI power. In Sec.~\ref{subsec:NLIMod}, we showed that memory is one of the main phenomena affecting the NLI. In previous NLI metrics (see Sec.~\ref{subsec:NLIMetric}), the effect of non-i.i.d. input sequences and their interaction with the channel memory was not considered. EDI brings together these two elements by capturing the statistical properties of the input sequence over a time window which is comparable to the channel memory. 

\subsection{Definition}\label{sec:EDI_Def}
EDI is designed to be a sliding window statistic with window length $W$. The windowed energy at time instant $i$, $G_{i}^{W}$, is defined as the total energy of $\boldsymbol{X}_{i-W/2}^{i+W/2}$, i.e.,
\begin{equation}\label{wndEngDef}
% \WE_{i}\triangleq\|\boldsymbol{X}^{W}_{i}\|^{2} = \sum_{j=i-W/2}^{i+W/2} |X_{j}|^2.
\WE_{i}\triangleq%\boldsymbol{\Energies}^{W}_{i} = 
\sum_{j=i-W/2}^{i+W/2} \Energy_{j}=\sum_{j=i-W/2}^{i+W/2} |X_j|^2.
\end{equation}  
%As shown in \eqref{wndEngDef}, the windowed energy is the sum of $W+1$ symbol energies centered around symbol $X_i$. 
Fig.~\ref{wndEngExmp} illustrates the windowed energies $\WE_{i}$ and $\WE_{i+1}$ and shows the sliding window effect.

The windowed energy is a random variable, and thus, we define the \emph{windowed energy process} as 
\begin{align}\label{WES}
[\ldots,\WE_{i-1},\WE_{i},\WE_{i+1},\ldots]. 
\end{align}
For CCDM QAM symbol sequences, the windowed energy process is the sum of $W+1$ time-shifted WSCS processes of symbol energies of period $n$, and thus the windowed energy process is also WSCS with the same period, i.e., $n$ \cite[Ch.~17.2-Prop.~1]{giannakis1998cyclostationary}. Therefore, $R_{\WE}(i,\tau)$ and $\Exp{\WE_i}$ are periodic with period $n$. Furthermore, the variance of the windowed energy $\WE_i$ (see \eqref{DefAutoCovStationary}) is given by
\begin{equation}\label{cycWndVar}
    \Var{\WE_i}=\Cov{\WE_i,\WE_i}=R_{\WE}(i,0)-\Exp{\WE_i}^2.
\end{equation}
With periodic $R_{\WE}(i,0)$ and $\Exp{\WE_i}$, $\Var{\WE_i}$ in \eqref{WES} also varies cyclically with period $n$.

% To evaluate the windowed energy variations, EDI is then defined as%\footnote{EDI has a similar definition as Fano factor, which was originally introduced to describe the strength of fluctuation for a Fano noise in particle detectors \cite{fano1947ionization}.}
% \begin{equation}\label{EDI}
% \Psi \triangleq\frac{\AvgVar{\WE}}{\AvgExp{\WE}},
% \end{equation}
% where the average windowed energy mean $\AvgExp{\WE}$ and the average windowed energy variance $\AvgVar{\WE}$ (similar to the definition of average autocorrelation in \eqref{AutoCorrDef}) are defined as
% \begin{align}
%  \AvgExp{\WE} &\triangleq \frac{1}{n} \sum_{i=0}^{n-1}  \Exp{\WE_i}, \label{AvgExpWe}
% \end{align}
% and
% \begin{align}
%  \AvgVar{\WE} &\triangleq  \frac{1}{n} \sum_{i=0}^{n-1} \Var{\WE_i}, \label{AvgVarWe}
% \end{align}
% respectively.
\begin{definition}[Energy Dispersion Index]\label{Def:EDI}
EDI is defined as
\begin{equation}\label{EDI}
\Psi \triangleq\frac{\AvgVar{\WE}}{\AvgExp{\WE}},
\end{equation}
where 
\begin{align}
 \AvgExp{\WE} &\triangleq \frac{1}{n} \sum_{i=0}^{n-1}  \Exp{\WE_i}, \label{AvgExpWe}\\
 \AvgVar{\WE} &\triangleq  \frac{1}{n} \sum_{i=0}^{n-1} \Var{\WE_i}. \label{AvgVarWe}
\end{align}
\end{definition}

EDI in Definition~\ref{Def:EDI} measures the windowed energy variations. EDI in \eqref{EDI} is defined as the ratio of the average windowed energy variance ($\AvgVar{\WE}$) to the average windowed energy mean ($\AvgExp{\WE}$). As shown in \eqref{AvgExpWe} and \eqref{AvgVarWe}, these two averages have the same form of the average autocorrelation $\AC$ in \eqref{AutoCorrDef}.

\subsection{Alternative Formulations}\label{sec:EDI_Fml}

EDI in Definition~\ref{Def:EDI} can also be expressed in terms of the second-order statistics of the input symbols. In this section, we introduce such formulation of the EDI for both CCDM QAM symbol sequences as well as i.i.d. symbol sequences. Note that we view the process of i.i.d. symbol sequences as a special case of a WSCS process with period $n=1$.

\begin{theorem}\label{Thm:WndVar} The average windowed energy mean and average windowed energy variance for CCDM QAM symbol sequences can be expressed as
\begin{align}\label{CCTExp}
    \AvgExp{\WE} &= (W+1)\Exp{|X|^2}
\end{align}
% and its  $\AvgVar{\WE}$ can be expressed as
and
\begin{align}
 \AvgVar{\WE}  &=  (W+1)\:\Var{|X|^2}-W(W+1)\Exp{|X|^2}^2 \nonumber \\
  &+ 2\sum_{\tau=1}^{W}(W-\tau+1)\AC, \label{wndTimeEngVar}
\end{align}
respectively, where $\AC$ is given by \eqref{AutoCorr}.
\end{theorem}

\begin{IEEEproof}
See Appendix~\ref{Proof:WndVar}.
\end{IEEEproof}

For i.i.d. symbol sequences, \eqref{CCTExp} also holds. On the other hand, $\Var{\WE_i}$ for i.i.d. symbol sequences is the sum of $W+1$ energy variances, therefore,
\begin{align}
 \AvgVar{\WE}  = &  \frac{1}{n} \sum_{i=0}^{n-1} \sum_{j=i-W/2}^{i+W/2} \Var{\Energy_j}\label{iidVar0} \\
 = & \frac{1}{n} \sum_{i=0}^{n-1} (W+1) \Var{|X|^2} \label{iidVar1} \\
 = & (W+1)\Var{|X|^2}, \label{iidVar}
\end{align}
where \eqref{iidVar1} follows from \eqref{VarStat}.

%In the next example, we study the behavior of $\AvgExp{\WE}$ and $\AvgVar{\WE}$ determining the EDI, as a function of the blocklength $n$.

\begin{figure}[!t]
\centering
% \resizebox{1\linewidth}{!}{\input{./Figures/wndEnergyPDF_histo.tikz}}
\includegraphics[width=1\linewidth]{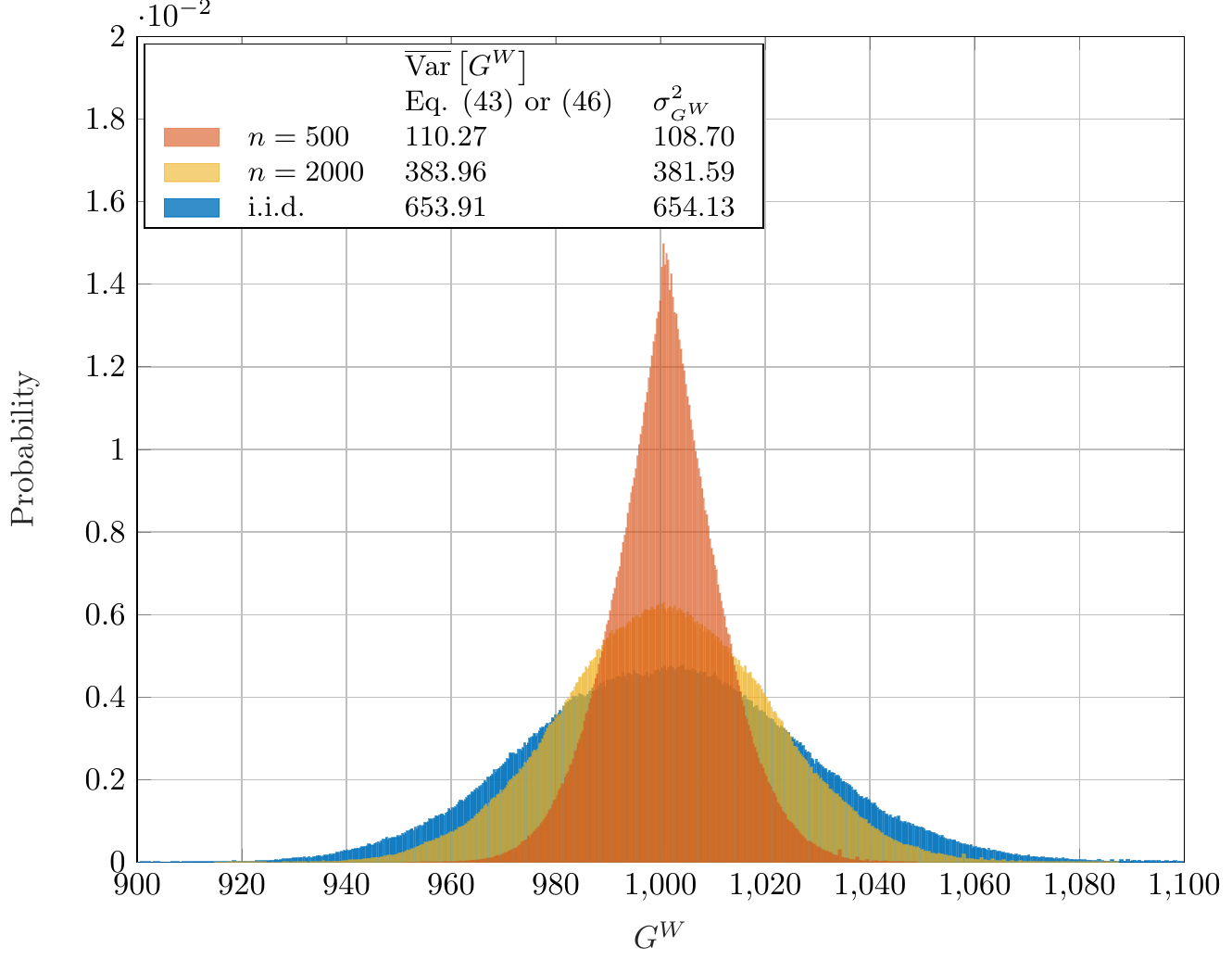}
\centering\caption{Histogram of windowed energy $\WE$ for $W=1000$. Normalized $\left(\Exp{|X|^2}=1\right)$ 64QAM symbol sequences with $\Prob_{A}=[0.4,0.3,0.2,0.1]$. The average windowed energy mean is $\AvgExp{\WE}=1000$.}
\label{PMF}
\end{figure}

\begin{example}[Windowed Energy Histograms]\label{Example4}
Fig.~\ref{PMF} illustrates how the probabilities of windowed energies depend on the blocklength $n$, based on a number of windowed energy samples. For simplicity, we only show the results of emulated CCDM amplitude codewords, since almost the same result is obtained when using exact CCDM. Fig.~\ref{PMF} shows that the mean value of $\WE_i$ is independent of $n$ as shown in \eqref{CCTExp}. Fig.~\ref{PMF} also shows that an increase of $n$ results in a heavier tail, i.e., a larger probability of observing a large windowed energy. The windowed energy variance increases as blocklength $n$ increases, which is in good agreement with our observations in Example~\ref{Example2}. Fig.~\ref{PMF} also shows that the estimated windowed energy variance $\TVar{\WE}$ is well-approximated by $\AvgVar{\WE}$ given in \eqref{wndTimeEngVar} and \eqref{iidVar}, where the discrepancies are caused by the limited number of samples.
\end{example}

Based on Theorem~\ref{Thm:WndVar}, the EDI of CCDM QAM symbol sequences can be obtained directly by substituting \eqref{wndTimeEngVar} and \eqref{CCTExp} in \eqref{EDI}, i.e., 
\begin{align}\label{CCDMEDI}
        \Psi(n,W) =\; & \Exp{|X|^2}\left[\Phi-(W+1)\right] \nonumber \\
        & + \frac{2\sum_{\tau=1}^W (W-\tau+1)\AC}{(W+1)\Exp{|X|^2}},
\end{align}
where $\AC$ and $\Phi$ are given by \eqref{AutoCorr} and \eqref{Kur}, respectively. Notation $\Psi(n,W)$ is used to emphasize the dependency of the EDI on the blocklength $n$ and window length $W$.

Apart from $n$ and $W$, in view of \eqref{Kur}, \eqref{bitcorr0} and \eqref{AutoCorr}, EDI in \eqref{CCDMEDI} is also determined by the kurtosis $\Phi$ as well as the second and fourth order moments of $|X|$. Furthermore, after dividing \eqref{iidVar} by \eqref{CCTExp} and using \eqref{Kur}, the EDI of i.i.d. is obtained, which is given by
\begin{equation}\label{iidEDI}
    \Psi=\Exp{|X|^2}(\Phi-1).
\end{equation}
It can be seen that \eqref{iidEDI} is independent of $n$ and $W$, but is still a function of the kurtosis $\Phi$. Recall that kurtosis from EGN model is derived based on i.i.d. symbols assumption, hence \eqref{iidEDI} means that EDI can give the same indication of the NLI as kurtosis does in the case of i.i.d. symbols.

We have derived a closed-form expression for the EDI in \eqref{CCDMEDI} (and for i.i.d. symbols in \eqref{iidEDI}). In the next section, we will investigate properties of the EDI, and compare them against the estimated EDI from Monte Carlo simulations.

\subsection{Properties}\label{sec:EDI_Prop}

In what follows, we first show that for certain values of blocklength $n$ and window length $W$, EDI depends linearly on $n$. This corresponds to a regime where $W$ (and thus, implicitly the channel memory) is larger than $n$. We then give bounds on EDI for arbitrary values of $n$ and $W$.

\begin{theorem}\label{Thm:LinEDI}
When $n\leq W+2$, the EDI $\Psi(n,W)$ in \eqref{CCDMEDI} depends linearly on $n$ via
\begin{equation}\label{CCDMEDI_Lin}
    \Psi_{\mathrm{lin}}(n,W) = \frac{n+1}{3(W+1)}\Exp{|X|^2}(\Phi-1).
\end{equation}
\end{theorem}
\begin{IEEEproof}
See Appendix~\ref{Appx:LinEDI}. 
\end{IEEEproof}
%It can be seen in \eqref{CCDMEDI_Lin} that for a fixed $W$, in the region of blocklength $n\leq W+2$, $\Psi(n,W)$ increases linearly with $n$.

\begin{corollary}\label{Thm:LowLmt}
For any finite blocklength $n$, the EDI in \eqref{CCDMEDI} is upper- and lower-bounded as
\begin{align}
    0 \leq \Psi(n,W) \leq \Exp{|X|^2}(\Phi-1).\label{EDIBounds}
\end{align}The bounds are achieved for asymptotic values of $W$, i.e.,
\begin{align}
\lim_{W \to 0} \Psi(n,W) &= \Exp{|X|^2}(\Phi-1).\label{upLimit} \\
\lim_{W \to \infty} \Psi(n,W) &= 0,\label{lowLimit}
\end{align} 
\end{corollary}
\begin{IEEEproof}
See Appendix~\ref{Appx:LowLmt}. 
\end{IEEEproof}

The lower bound in \eqref{lowLimit} can be intuitively understood as follows. When $W$ is much larger than $n$, the windows always include multiple complete blocks of QAM symbols, and the compositions of amplitudes within a large window ``hardens", yielding a reduced fluctuation of the window energies. The upper bound in \eqref{upLimit} is identical as the EDI of i.i.d. symbol sequences in \eqref{iidEDI}. This upper bound indicates that as window length decreases (less memory in the metric), the impact of symbol energy correlations on EDI decreases.%, and thus, exhibits the same EDI as that of i.i.d. symbol sequences.

Note that the EDI for constant-modulus constellations (such as, e.g., phase-shift keying), is identically 0, due to the fact that the symbols have constant energy. The sliding window energy is thus constant. This reflects the fact that EDI is a metric specifically designed to capture NLI fluctuations in PAS systems with different shaping blocklengths, and such systems cannot be designed using constant modulus constellations.

% EDI generally quantifies the windowed energy fluctuation over time. To estimate the EDI in Monte Carlo simulations, a finite number of windowed energy samples $\WES=\boldsymbol{g}^W$ is constructed based on a realization of symbol sequence $\boldsymbol{X}=\boldsymbol{x}$. The first and last windowed energy samples are obtained when the window slides to the edge of the symbol sequence. With the estimated windowed energy variance $\TVar{\WE}$ and the estimated windowed energy mean $\TExp{\WE}$, the estimated EDI is
% \begin{equation}
% \label{EDIEst}
% \hat{\Psi} \triangleq \frac{\TVar{\WE}}{\TExp{\WE}}.
% \end{equation} 

\begin{figure}[!t]
\centering
% \resizebox{1\linewidth}{!}{\input{./Figures/DvsN.tikz}}
\includegraphics[width=1\linewidth]{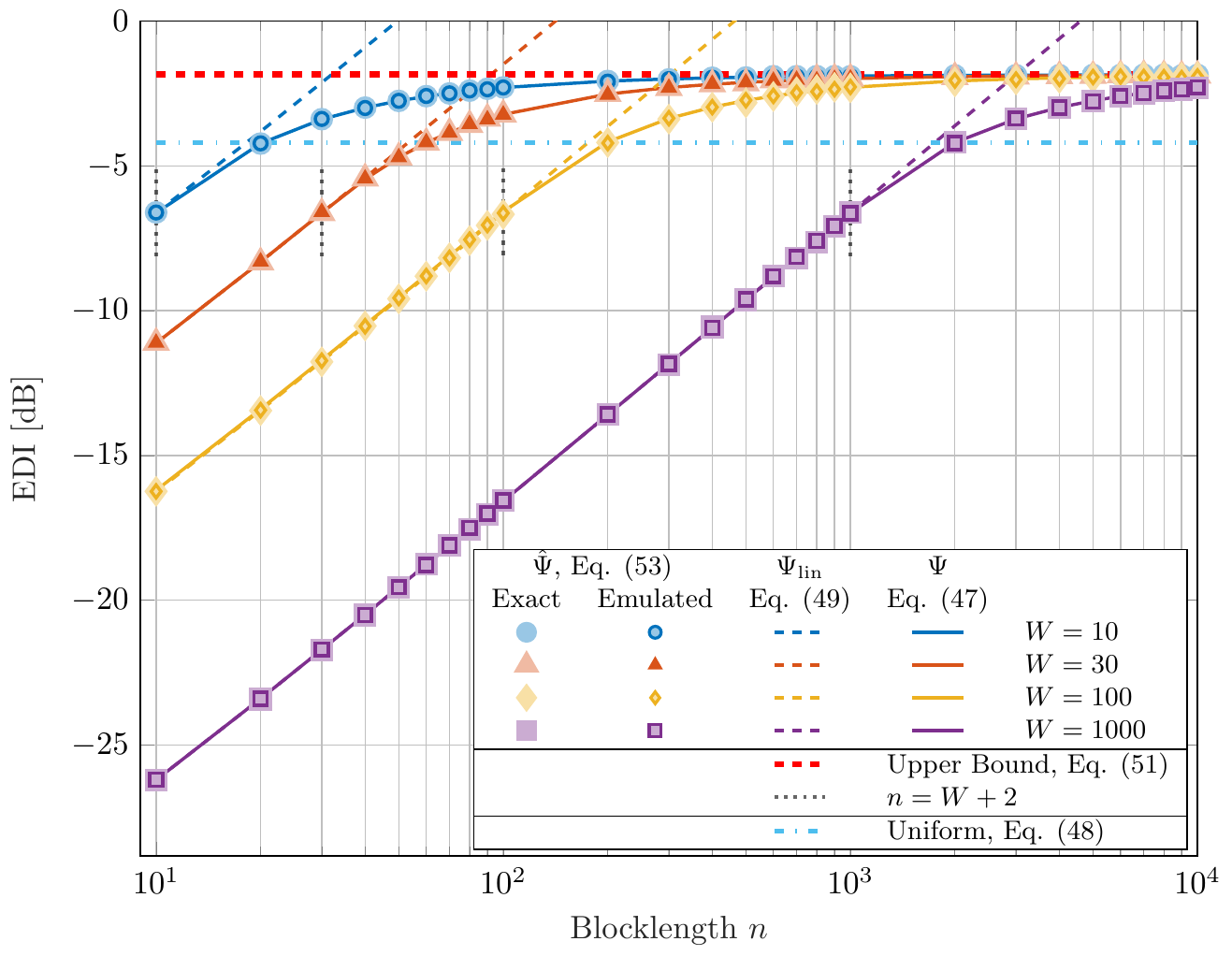}
\centering\caption{Simulation results (markers) using \eqref{EDIEst}, analytical results (solid lines) in \eqref{CCDMEDI} and \eqref{iidEDI}, and linear EDI in \eqref{CCDMEDI_Lin} (in dB) vs. blocklength for normalized $\left(\Exp{|X|^2}=1\right)$ 64QAM symbol sequences with $\Prob_{A}=[0.4,0.3,0.2,0.1]$.}
\label{DvsN}
\end{figure}

\begin{example}[EDI and Blocklength]\label{Example.EDI.n}
Fig.~\ref{DvsN} shows the EDI (in dB) of symbol sequences for different values of $n$ and $W$, as well as the EDI for i.i.d. uniform 64QAM sequences. The analytical results of EDI are computed by using \eqref{CCDMEDI}, \eqref{iidEDI} and \eqref{CCDMEDI_Lin} as well as the EDI estimated as
\begin{equation}
\label{EDIEst}
\hat{\Psi} \triangleq \frac{\TVar{\WE}}{\TExp{\WE}},
\end{equation} 
where $\TVar{\WE}$ and $\TExp{\WE}$ are the estimated windowed energy variance and the estimated windowed energy mean, respectively. Fig.~\ref{DvsN} shows that all the simulation results $\hat{\Psi}$ match the analytical results $\Psi$ in \eqref{CCDMEDI} very well. This is the case even for short blocklength ($n=10$), indicating that the CCDM emulation approach we took in this paper has little impact on the EDI in \eqref{CCDMEDI}. Fig.~\ref{DvsN} also shows the linear EDI expression $\Psi_{\mathrm{lin}}$ in \eqref{CCDMEDI_Lin}. The EDI curves can be seen to be segmented into a linearly blocklength-dependent region by $n=W+2$. As $n$ increases above $W+2$, $\Psi$ begins to diverge from $\Psi_{\mathrm{lin}}$ in \eqref{CCDMEDI_Lin}. For a fixed $n$, as $W$ increases, EDI approaches the lower bound in \eqref{lowLimit} (i.e., $-\infty$ in dB). On the other hand, EDI gradually reaches the upper bound in \eqref{upLimit} as $W\to 0$.
\end{example}

It can be concluded from Example~\ref{Example.EDI.n} that for CCDM QAM symbol sequences, if a symbol-level interleaver, or a very long blocklength is used, the EDI will approach the upper bound in \eqref{upLimit} that is determined by the kurtosis. Hence, EDI can be viewed as a windowed version of kurtosis. For the purpose of NLI prediction, window length $W$ should be carefully chosen such that the channel memory effect is properly reflected, as will be shown in the following section. 

We conclude this section by emphasizing that Example~\ref{Example.EDI.n} (and also  Example~\ref{Example3}) showed that the CCDM emulation approach of considering all $N_C$ permutations as codewords gives very precise results. This assumption allowed us to find closed-form expressions for the average autocorrelation (see \eqref{AutoCorr}) and the EDI (see \eqref{CCDMEDI}). In the next section, we therefore show results using the analytical expressions we have developed above, and thus, only consider emulated CCDM.

\section{Numerical Results}\label{sec:NumAna}

In this section, we show that EDI can qualitatively predict the NLI magnitude when different blocklengths are used. To this end, we study the effective SNR in \eqref{effSNR0}, where NLI is a substantial part of the total noise at relatively high power, and where NLI changes produce a change in effective SNR.

\subsection{Simulation Setup}

We consider an ideal single-polarization multi-span WDM fiber system with $N_{\mathrm{ch}}=5$ channels. Nonlinear noise caused by the Kerr effect, as well as ASE and CD are taken into consideration. The fiber propagation is simulated using the split-step Fourier method with a step size of $50$~m. Other key simulation parameters are displayed in Table~\ref{FiberParam}. The channel of interest is located at the center of the WDM spectrum, where the channel spacing is $\Delta f=50$~GHz. The signal is generated with root-raised cosine pulse shaping. After propagation over each span of standard single-mode fiber with span length 80 km, the attenuation is ideally compensated by an Erbium-doped fiber amplifier (EDFA). At the receiver, the channel of interest is filtered with a matched filter, followed by CD compensation and sampling.

%% Parameter table
\begin{table}[!t]
    \centering
    \caption{Simulation Parameters.}
    \label{FiberParam}
    \begin{tabular}{c|c} 
    \hline\hline\textbf{Parameter} & \textbf{Value} \\
    \hline Modulation & $64$ QAM \\
    Pol.-mux. & Single \\
    Wavelength ($\lambda$) &  1550 nm \\
    Symbol rate & $32$ GBd \\
    WDM spacing ($\Delta f$) & $50$ GHz \\
    WDM channels ($N_{\mathrm{ch}}$) & 5 \\
    Pulse shape & root-raised cosine \\
    Roll-off & $10\% $ \\
    \hline Fiber length & $80$ km\\
    %No. of spans & $1$, $4$ and $10$ \\
    Fiber loss & $0.2$ dB/km \\
    Dispersion parameter ($D$) & $17$ ps/nm/km \\
    Nonlinear parameter ($\gamma$) & $1.37$ 1/W/km \\
    %Power per ch. & $-7$ to $1.5$ dBm \\
    EDFA noise figure & $6$ dB \\
    \hline Oversampling factor & $2 \times$ \\
    QAM symbols per run & $2^{18}$ \\
    No. of simulation runs & 10 \\\hline\hline
    \end{tabular}
\end{table}

The \emph{one-sided} channel memory $M$ is an important reference for the window length $W$ in the EDI. $M$ can be estimated using \eqref{ChnMm1} and \eqref{ChnMm2}. In \eqref{ChnMm2}, the optical bandwidth is $\Delta\omega=N_{\mathrm{ch}}\Delta f$, while group velocity dispersion $\beta_{2}$ is related to dispersion parameter $D$ as shown in \cite[Eq.~(1.2.11)]{agrawal2013nonlinear}. 

For shaped 64QAM transmission, the amplitude PMF $\Prob_{A}=[0.4,0.3,0.2,0.1]$ is used. Note that EDI can be used for arbitrary PMF in principle. Quadrature phase-shift keying (QPSK) and uniform 64QAM are presented as baselines. QPSK is anticipated to provide optimal effective SNR performance, since constant modulus constellations completely remove the modulation-dependent NLI \cite{dar2014shaping}. We also consider shaped 64QAM symbol sequences using a randomly generated symbol-level interleaver to emulate i.i.d. symbol sequences. The symbol sequences are always normalized to $\Exp{|X|^2}=1$. 

%% Signaling Metric Table
\begin{table}[!t]
    \centering
    \caption{NLI Metrics of the Evaluated Symbol Sequences. \protect\\ The results for $\text{R}_r$ are taken from \protect\cite[Fig.~8]{fehenberger2019analysis}.}
    \label{MetricTable}
    \begin{tabular}{m{1.7cm}|m{1.0cm} m{1.0cm} m{1.0cm} m{1.0cm}}
    \hline\hline\textbf{Metrics} & CC PS 64QAM & i.i.d. PS 64QAM & Uniform 64QAM & QPSK \\\hline
     $\Theta$, PAPR & $3.769$ & $3.769$ & $2.336$ & $1$\\
     $\Phi$, Kurtosis & $1.653$ & $1.653$ & $1.381$ & $1$\\
     $\text{R}_r$, Run Ratio& $\approx 0.98$ & $0.978$ & $0.984$ & $0.750$\\
     %$\Psi$, EDI & Eq.~\eqref{CCDMEDI} & $0.653$ & $0.381$ & $0$\\
     \hline\hline
    \end{tabular}
\end{table}

\begin{figure}[!t]
  \begin{minipage}{1\linewidth}
    \centering
    % \resizebox{1\linewidth}{!}{\input{./Figures/SNR_DvsN_80km.tikz}}
    % \resizebox{1\linewidth}{!}{\input{./Figures/SNR_DvsN_320km.tikz}}
    % \resizebox{1\linewidth}{!}{\input{./Figures/SNR_DvsN_1600km.tikz}}
    \includegraphics[width=1\linewidth]{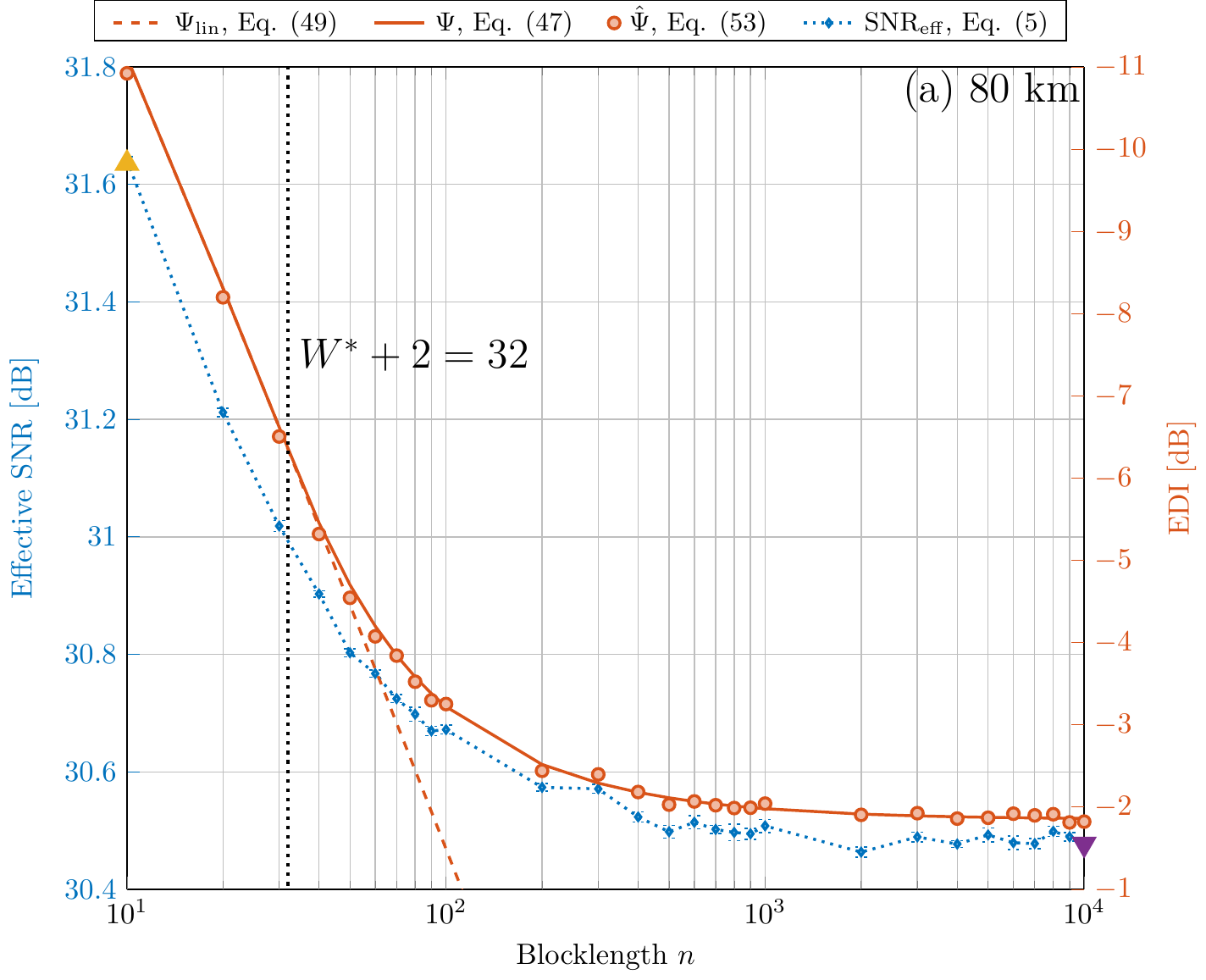}
    \includegraphics[width=1\linewidth]{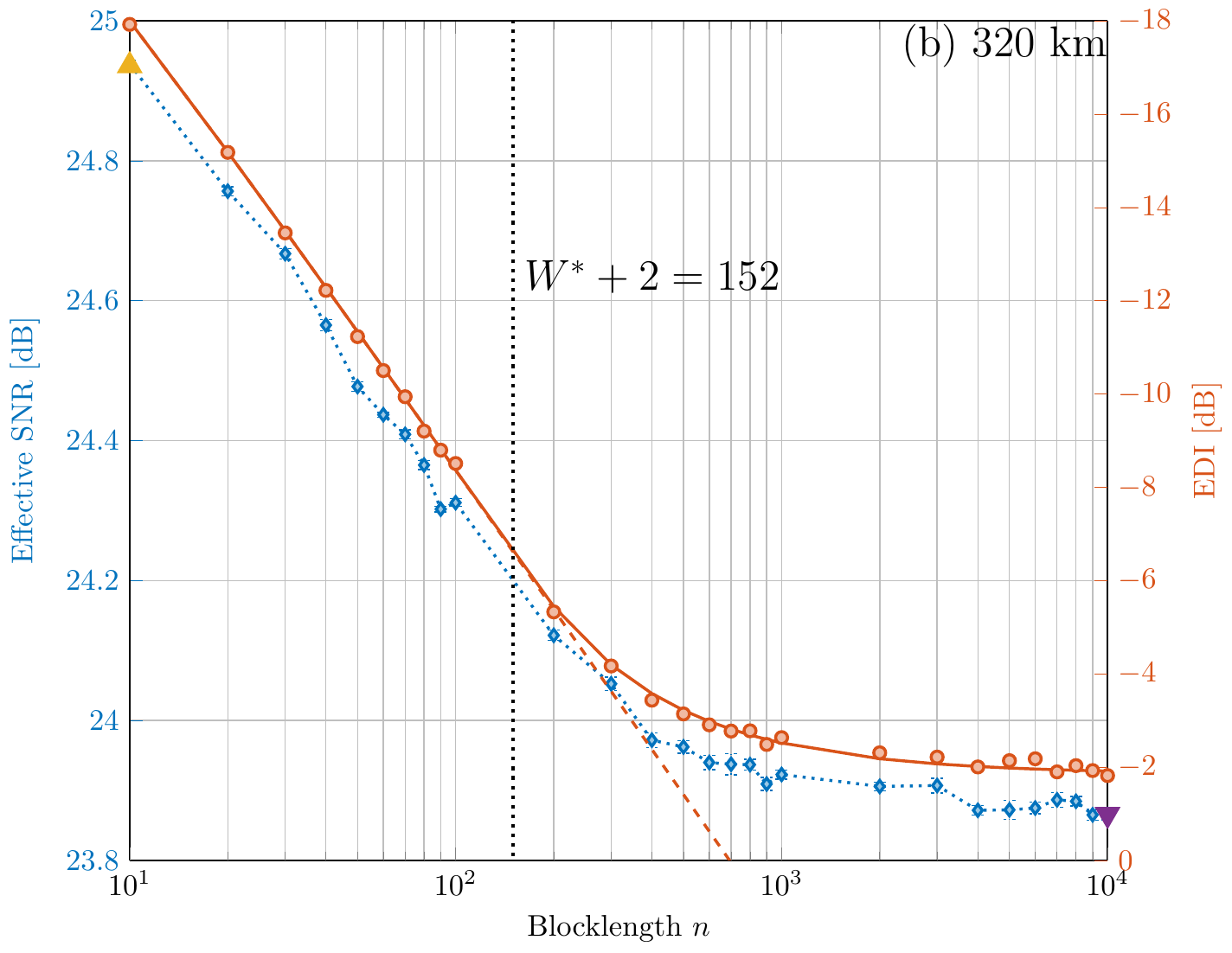}
    \includegraphics[width=1\linewidth]{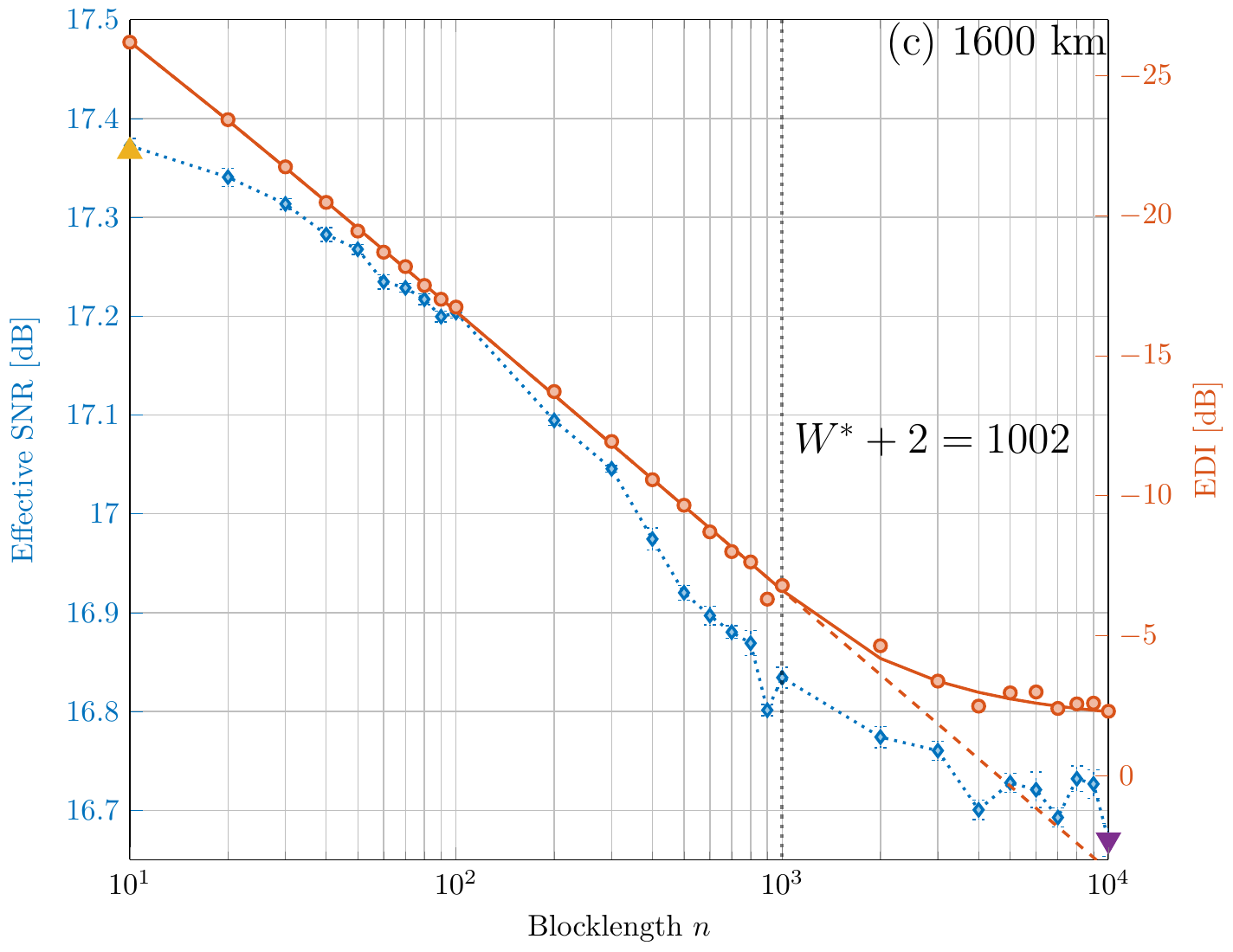}
    \caption{Effective SNR (left axis) and EDI (right axis) vs. blocklength after transmission distances of (a) 80 km, (b) 320 km and (c) 1600 km. The launch powers are (a) $-1.5$ dBm, (b) $-2.0$ dBm and (c) $-3.0$ dBm. Error bars for effective SNRs represent 95\% confidence interval. The EDI is shown in dB and inverted for convenience of comparison. The optimal window lengths $W^{*}$ shown in the figure are used for the EDI calculation. The correlation coefficients $r_p$ in \eqref{PsCorr} are (a) $-0.998$, (b) $-0.999$ and (c) $-0.995$.}
    \label{SNRD}
  \end{minipage}
\end{figure}

\begin{figure}[!t]
\centering
% \resizebox{1\linewidth}{!}{\input{./Figures/CorrvsWnd.tikz}}
\includegraphics[width=1\linewidth]{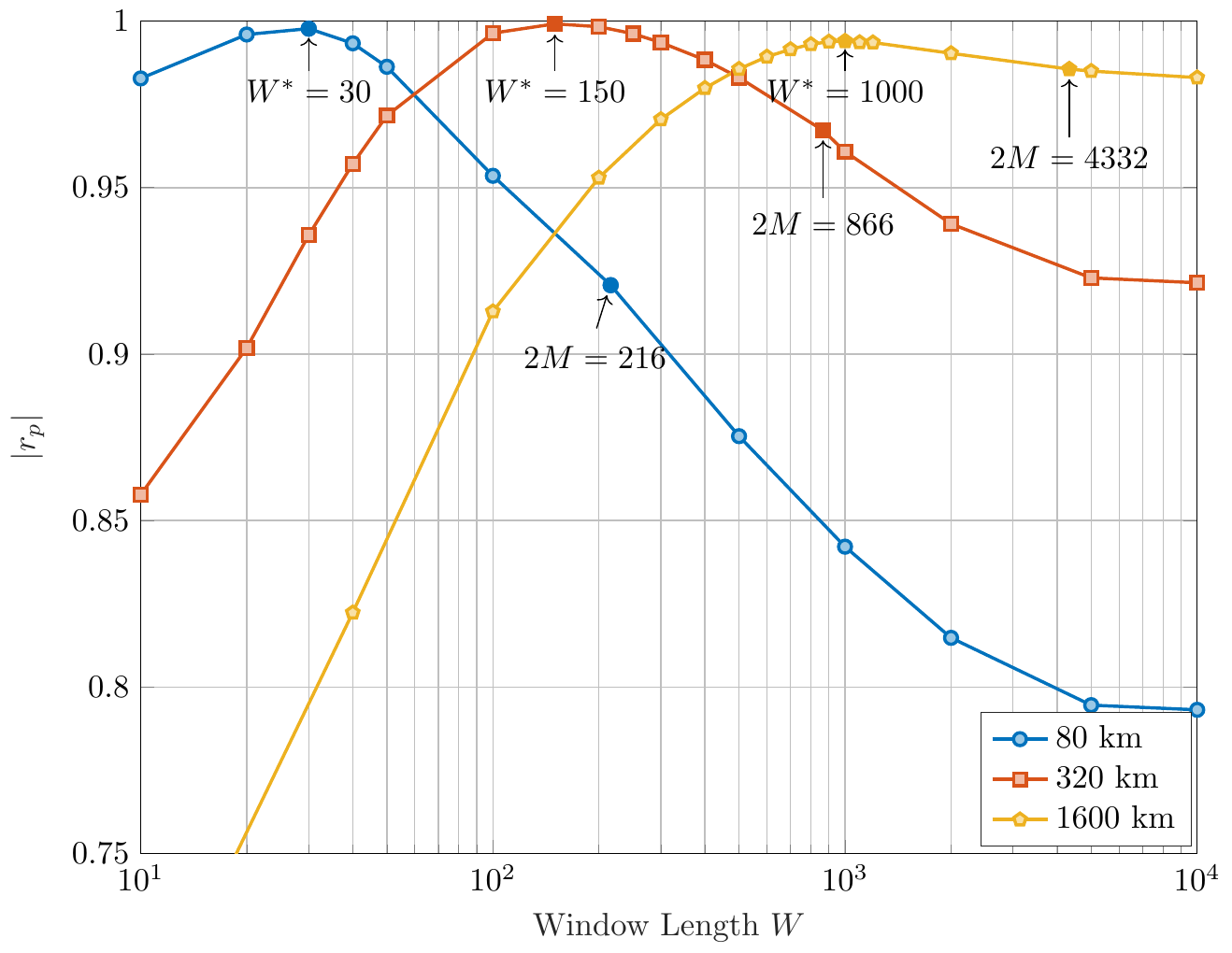}
\centering\caption{Absolute value of Pearson's linear correlation coefficient $|r_p|$ in \eqref{PsCorr} between EDI and effective SNR vs. window length $W$ for $80$ km, $320$ km and $1600$ km, whose launch powers are $-1.5$ dBm, $-2.0$ dBm and $-3.0$ dBm, respectively. The optimal value of $W$ is denoted by $W^{*}$. The channel memories $2M$ calculated using \eqref{ChnMm1}--\eqref{ChnMm2} are also indicated.}
\label{CorrvsW}
\end{figure}

An overview of the NLI-related metrics is given in Table~\ref{MetricTable}. PAPR and kurtosis do not depend on the blocklength, and thus, these two metrics cannot predict the NLI differences caused by blocklength differences. Although run ratio can, to some extent, capture the blocklength dependency of the NLI, it does not take channel memory into account, and thus, cannot adapt to different distances. By contrast, EDI uses a window length depending on the distance.

With various blocklengths $n$, many pairs of effective SNR and EDI can be obtained. To quantify the correlation between effective SNR and EDI, Pearson's correlation coefficient \cite[Ch.~11.1]{gibbons2014nonparametric} is used, i.e.,
\begin{equation}\label{PsCorr}
    r_p \triangleq \frac{\mathrm{Cov}\bigl[\SNRe,\hat{\Psi}\bigr]}{\sqrt{\Var{\SNRe}\mathrm{Var}\bigl[\hat{\Psi}\bigr]}}.
\end{equation}
Coefficient values $+1$ or $-1$ indicates perfect correlation, while $0$ indicates no correlation.

\subsection{EDI and Effective SNR}
We first investigate how EDI is correlated to effective SNR. Fig.~\ref{SNRD} displays effective SNR and EDI vs. blocklength. At each transmission distance, the optimal launch power found for $n=10$ is used (see Fig.~\ref{SNRP} ahead). The effective SNRs achieved for $n=10$ and $n=10000$ at these launch powers are shown with filled triangles in Fig.~\ref{SNRD}. EDI is calculated using the optimal window lengths $W^{*}$, which will be discussed in Fig.~\ref{CorrvsW}. Fig.~\ref{SNRD} shows that the estimated EDIs (circles) are in good agreement with the analytical EDI in \eqref{CCDMEDI} (solid lines), despite of slight fluctuations due to the limited number of transmitted symbols. Effective SNR almost follows the same trend as that of EDI, and their Pearson's linear correlation coefficients at three distances are at least $99.5\%$. Based on Theorem~\ref{Thm:LinEDI}, the x-axes in Fig.~\ref{SNRD} can be divided into two regions: $n<W^{*}+2$ and $n>W^{*}+2$. The blocklength-dependent region on the left shows that the effective SNR varies linearly with blocklength $n$. As $n$ increases entering the region on the right, SNR begins to decreases slowly until exhibiting marginal differences, since for long blocklengths the NLI reduction brought by weakly-correlated symbol energies becomes insignificant. In this region, EDI is determined by kurtosis, and the EGN model is able to give accurate predictions of effective SNR, as demonstrated in \cite[Fig.~7]{fehenberger2019analysis}.
 
The optimal window length $W^{*}$ used in Fig.~\ref{SNRD} was chosen such that EDI yields the highest correlation with effective SNR. To this end, $W^{*}$ is obtained by analyzing various window length $W$ at each transmission distance, whose absolute value of correlation coefficient $|r_p|$ is shown in Fig.~\ref{CorrvsW}. It can be seen that $|r_p|$ reaches its peak for values $W^{*}$, which is much smaller than the estimated channel memory $2M$ given by \eqref{ChnMm1}--\eqref{ChnMm2}. Incorrectly choosing $W$ can lead to smaller values of $|r_p|$, and thus the SNR prediction by EDI is less accurate. However, Fig.~\ref{CorrvsW} also shows that even if $2M$ is used as window length, EDI can still reflect the SNR variations with correlation coefficients above $90\%$. Thus, in practice one can directly use the estimated channel memory $2M$ as window length rather than finding the optimal one $W^{*}$. In general, Fig.~\ref{SNRD} and Fig.~\ref{CorrvsW} show that EDI and effective SNR are highly correlated with each other, indicated by a nearly perfect negative correlation. It can be concluded that EDI evaluated with the optimal window length is capable of reflecting the impact of blocklength and distance on the NLI.

\begin{figure}[!t]
\centering
% \resizebox{1\linewidth}{!}{\input{./Figures/optWndvsDst.tikz}}
\includegraphics[width=1\linewidth]{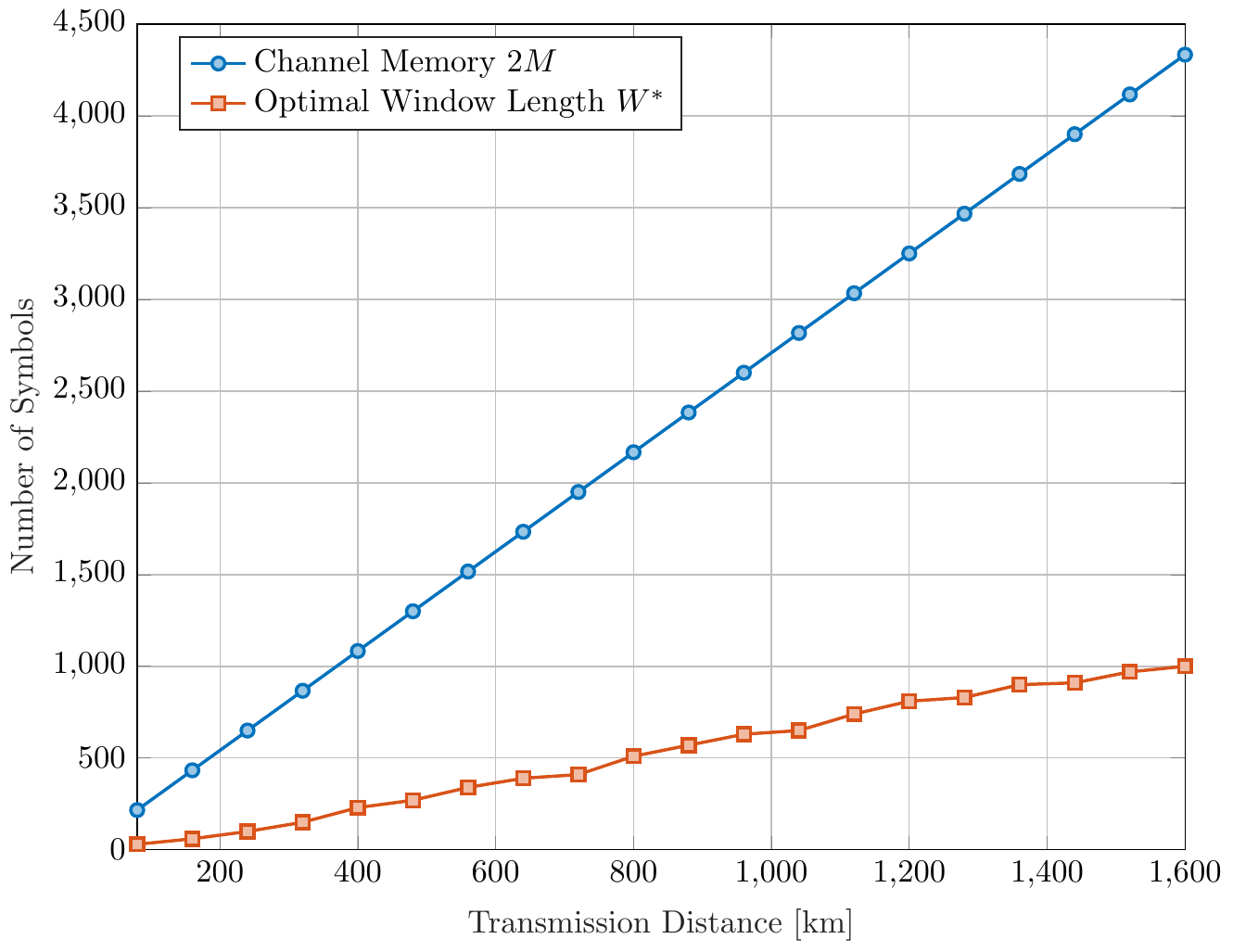}
\centering\caption{The estimated channel memory $2M$ calculated using \eqref{ChnMm1}--\eqref{ChnMm2} and the optimal window length $W^{*}$ ($|r_p|\geq0.995$) at transmission distances from $80$ km to $1600$ km.}
\label{optWvsDst}
\end{figure}

\begin{figure}[!t]
  \centering
  \begin{minipage}{1\linewidth}
    \centering
    % \resizebox{1\linewidth}{!}{\input{./Figures/SNRvsPwr_80km.tikz}}
    % \resizebox{1\linewidth}{!}{\input{./Figures/SNRvsPwr_320km.tikz}}
    % \resizebox{1\linewidth}{!}{\input{./Figures/SNRvsPwr_1600km.tikz}}
    \includegraphics[width=0.95\linewidth]{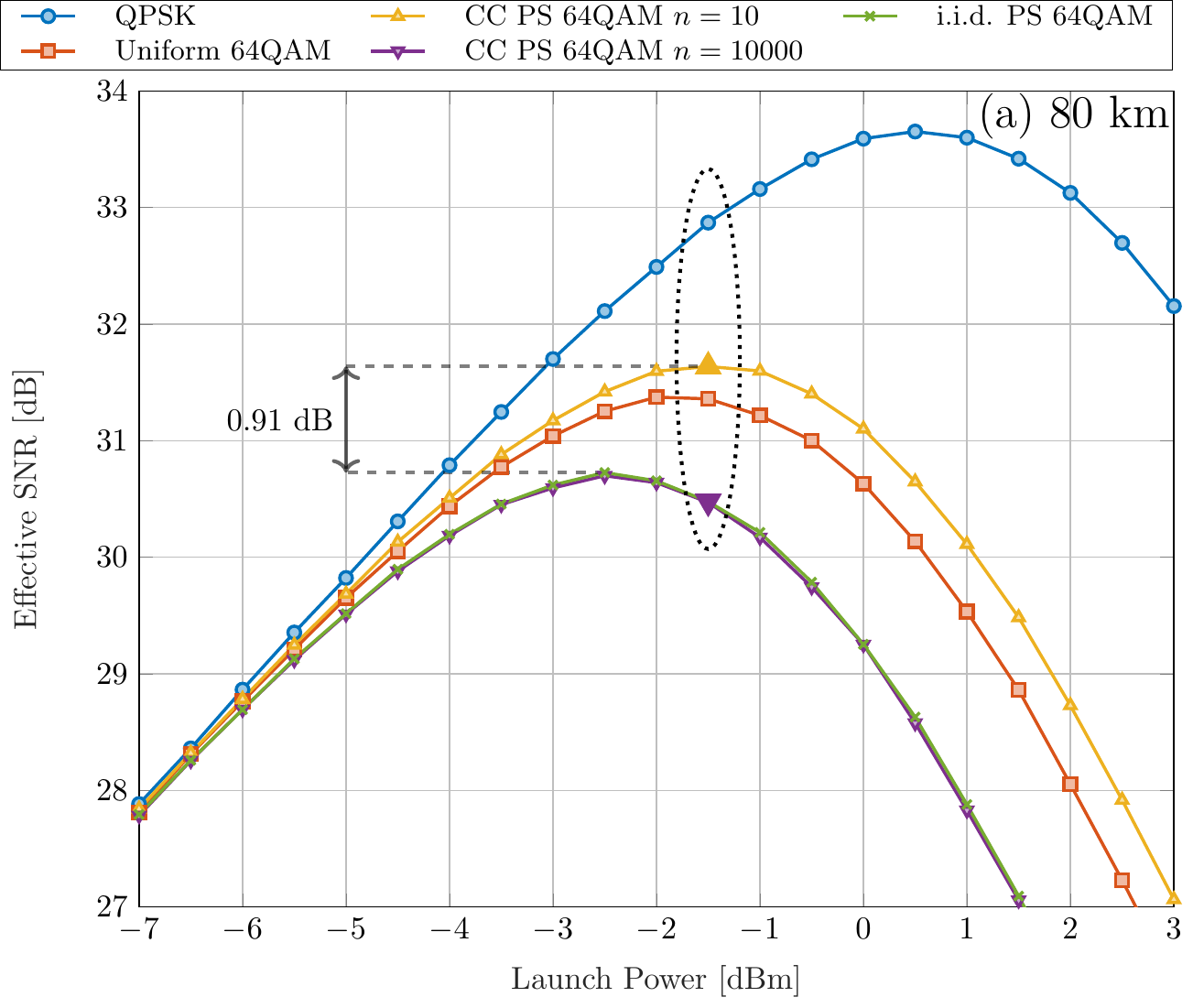}
    \includegraphics[width=0.95\linewidth]{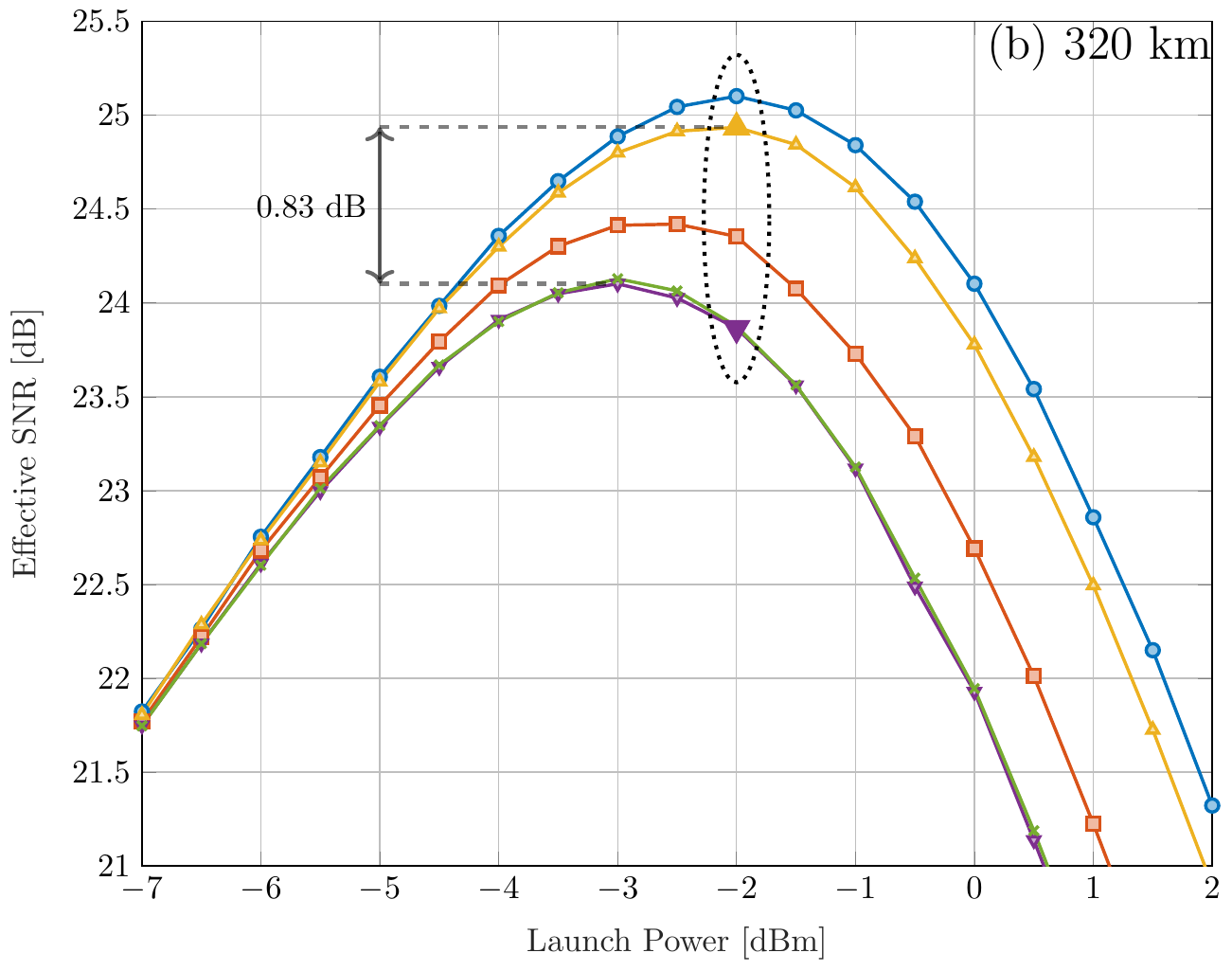}
    \includegraphics[width=0.95\linewidth]{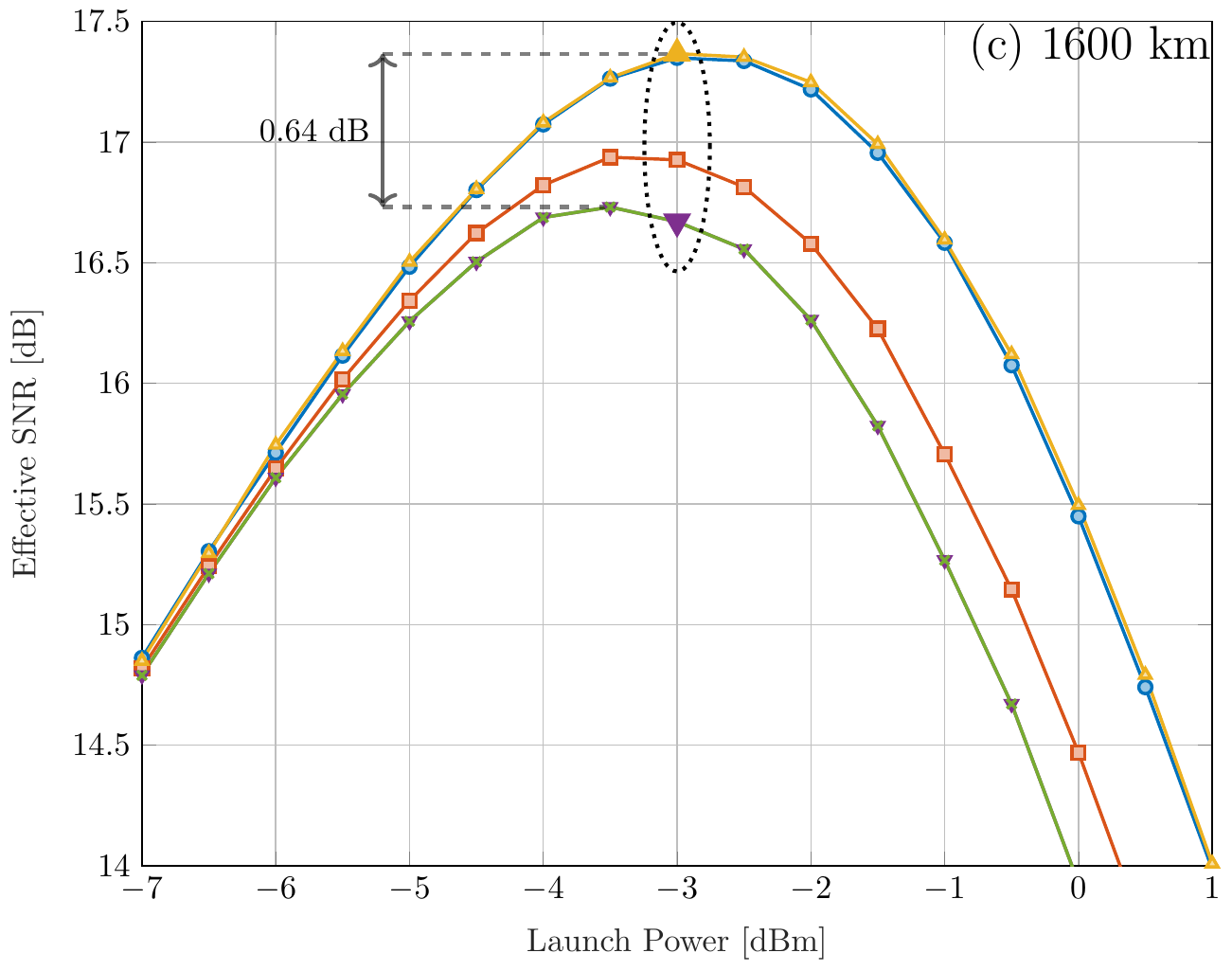}
    \caption{Effective SNR vs. launch power after transmission distances of (a) 80 km, (b) 320 km and (c) 1600 km. PAS 64QAM with blocklength $n=10$ and $n=10000$ are displayed. Results of uniform 64QAM, i.i.d. PS 64QAM symbols, and QPSK are also included as references. The circled SNR performance correspond to the launch powers used in Fig.~\ref{SNRD}~(a)--(c) respectively.}
    \label{SNRP}
  \end{minipage}
\end{figure}

%% Signaling Metric Table
\begin{table}[!t]
    \centering
    \caption{EDI $\Psi$ (in dB) of the Evaluated Symbol Sequences}
    \label{EDITable}
    \begin{tabular}{m{1cm}|ccccc}
    \hline\hline
    \multirow{2}*{\textbf{Distance}} & \multicolumn{2}{c}{CC PS 64QAM} & i.i.d. PS & Uniform & \\
    & $n=10$ & $n=10000$ & 64QAM & 64QAM & \multirow{-2}*{QPSK} \\
    \hline
    80 km & $-11.12$ & $-1.86$ & $-1.85$ & $-4.20$ & $-\infty$\\
    320 km & $-18.00$ & $-1.91$ & $-1.85$ & $-4.20$ & $-\infty$\\
    1600 km & $-26.21$ & $-2.29$ & $-1.85$ & $-4.20$ & $-\infty$\\
     \hline\hline
    \end{tabular}
\end{table}

Fig.~\ref{optWvsDst} shows the relationship between $W^{*}$ and estimated channel memory $2M$ at various transmission distances. All $W^{*}$ are obtained with at least $99.4\%$ absolute correlation coefficients $|r_p|$. It can be seen in Fig.~\ref{optWvsDst} that $M$ scales linearly with distance, as $M$ is computed by using \eqref{ChnMm2}. Likewise, $W^{*}$ also increases approximately linearly but at a slower rate. The optimal window length $W^{*}$ is smaller than the estimated number of interfering symbols $2M$, which can be explained by two facts. First, \eqref{ChnMm2} is a rough estimation of the channel memory. Second, the symbol energies within EDI window are assumed to have equal contributions to the NLI (as shown in \eqref{wndEngDef} that all symbol energies are weighted by $1$). Therefore, $W^{*}$ generally represents an effective number of dominant interfering symbols that are involved in the NLI generation.

To conclude, we show effective SNR vs. launch power in Fig.~\ref{SNRP}, which shows that the impact of the shaping blocklength on effective SNR can be as large as that of modulation format. For simplicity, CCDM 64QAM symbol sequences are only shown with ultra short blocklength $n=10$ and long blocklength $n=10000$, in general representing ``good" and ``bad" NLI-tolerant blocklengths. The EDI of different symbol sequences are given in Table~\ref{EDITable}. The filled triangles in Fig.~\ref{SNRP} for $n=10$ and $n=10000$ correspond to the same effective SNR markers as shown in Fig.~\ref{SNRD}. The first observation from Fig.~\ref{SNRP}~(a)--(c) is that QPSK exhibits the best effective SNR for the distances under consideration, as it has the smallest EDI. Secondly, the effective SNRs of i.i.d. 64QAM symbol sequences and the shaped symbol sequence with $n=10000$ almost coincide for all distances. Meanwhile, their EDIs have marginal differences. Thirdly, the effective SNR of uniform 64QAM falls between the effective SNRs of shaped 64QAM using $n=10$ and $n=10000$, so does their EDIs. Lastly, Fig.~\ref{SNRP} shows that the SNR gains offered by CCDM symbol sequences using $n=10$ instead of $n=10000$ decrease from $0.91$~dB to $0.64$~dB as transmission distance increases. % In Fig.~\ref{SNRP}~(c), shaped 64QAM with $n=10$ even exhibits slightly higher SNR performance than QPSK, which contradicts the prediction of EDI. One explanation is that EDI only captures the NLI reductions aroused by the dependency between the second-order moments (energies) of symbols, while it neglects the NLI reductions caused by the dependency between higher-order moments. Moreover, this observation implies the possibility that QAM modulation formats can suppress even more amount of NLI as standard QPSK transmission through transmitting correlated symbols. 

%%%%  Conclusion  %%%%%%%%%%%%%%%%%%%%%%%%%%
\section{Conclusions}\label{sec:Conc}

This paper proposed a new heuristic metric called energy dispersion index (EDI) to predict the impact of blocklength on the effective SNR for CCDM-coded systems. EDI is a measure of the windowed energy dispersion which captures the interaction between the statistical properties of a CCDM input sequence and the channel memory with respect to the received NLI magnitude. Numerical results show that the effective SNR is highly correlated to the EDI of the transmitted symbol sequence, with correlation coefficients greater than $99\%$.

Being a heuristic metric, the EDI requires future analytical substantiation, possibly leading to a refined version thereof. One possible improvement of the EDI accuracy could be appropriately weighting the symbol energies within the EDI window to reflect their uneven contributions to the NLI. %In this paper, we investigated EDI of symbol sequences in the time domain, however, more insights by analyzing EDI in the frequency domain could be obtained. 
In this paper, we only studied EDI for CCDM, and thus, a performance analysis of EDI to other shaping algorithms or other constellations is still pending. All these open problems are left for future work. Nevertheless, we believe that EDI can facilitate the development of NLI-tolerant signaling schemes that aim to optimize time-varying statistical properties of the input symbol sequence.

%%%%  Appendix  %%%%%%%%%%%%%%%%%%%%%%%%%%%%%%%%%%
\section*{Acknowledgments}
The authors would like to thank Dr. Yunus Can G\"ultekin and Sebastiaan Goossens (Eindhoven University of Technology) for fruitful discussions on shaping techniques.

\appendices

\section{Proof of Theorem~\ref{Thm:NegCov}}\label{Proof:NegCov}
In this Appendix, we start by computing $R_{\Energy}(i,\tau)$ in \eqref{autoCorrEnergy}. For $i=0,1,\ldots,n-1$ (one period), $R_{\Energy}(i,\tau)$ is given by
\begin{equation}\label{AutoCorrSplt}
\setlength{\nulldelimiterspace}{0pt}
R_{\Energy}(i,\tau) =\left\{\begin{IEEEeqnarraybox}[\relax][c]{l's}
\Exp{|X|^4}, &if $\tau=0$\\
\CrS, &if $\tau\neq 0, 0 < i+\tau < n $\\
\Exp{|X|^2}^2, &if $\tau\neq 0, i+\tau \geq n $ or $i+\tau \leq 0 $
\end{IEEEeqnarraybox}\right.
\end{equation}

The second case in \eqref{AutoCorrSplt} is when two different symbol energies belong to the same block. In what follows we will prove that in this case symbol energies are equally correlated and yields the same autocorrelation $\CrS$ in \eqref{bitcorr0}, and also autocovariance. By using Lemma~\ref{Thmmin1}, \eqref{AutoCorrDef0} and \eqref{DefAutoCovStationary},  for $\forall i,j\in\{0,1,\ldots,n-1\}$ and $i\neq j,\tau =j-i$, we can write
\begin{align}
    \Cov{\Energy_i,\Energy_j} = & R_{\Energy}(i,j-i)-\Exp{\Energy}^2 \label{CovAmp0}\\
     = & \Exp{(A_{I,i}^2+A_{Q,i}^2)(A_{I,j}^2+A_{Q,j}^2)}-4\Exp{A^2}^2 \label{CovAmp1}\\
     = & 2\Exp{A_i^2A_j^2}-2\Exp{A^2}^2, \label{CovAmp2}\\
     = & 2\sum_{a,b\in\mathcal{A}}\Prob_{A_i,A_j}(a,b)a^2 b^2-2\Exp{A^2}^2,\label{CovAmp3}
\end{align}
where \eqref{CovAmp1} follows from \eqref{EnergyDef}, \eqref{EX2Stat} and \eqref{EX2}, and \eqref{CovAmp2} from the fact that amplitudes in the I/Q branches in Fig.~\ref{PASsys} are independent from each other. It can be seen in \eqref{CovAmp3} that $R_{\Energy}(i,j-i)$ and $\Cov{\Energy_i,\Energy_{j}}$ is determined by the joint probability $\Prob_{A_i,A_j}$, i.e.,
\begin{align}\label{jointProb}
\Prob_{A_i,A_j}(a,b)=&\Prob_{A_i}(a)\cdot\Prob_{A_j|A_i}(b|a),
\end{align}
where $\Prob_{A_i}(a)=n_a/n$ (see \eqref{A.stationary}). For a given $A_i=a$, the amplitude composition at time $j$ is updated, and thus,
\begin{equation}\label{ProbSplt}
\Prob_{A_j|A_i}(b|a)=\left\{\begin{array}{lr}
\frac{n_a-1}{n-1}, & \text{if}\; b=a\\
\frac{n_b}{n-1}, &\text{if}\; b\neq a
\end{array}.\right.
\end{equation} 
The joint probability $\Prob_{A_i,A_j}$ in \eqref{jointProb} is thus, independent of $i$ and $j$, and so is $R_{\Energy}(i,j-i)=\CrS$ and $\Cov{\Energy_i,\Energy_{j}}$. 

Finally, based on the fact that $\CrS$ is a constant, we compute $\CrS$ and the autocovariance. Similar to \eqref{linear}, the total energy of a symbol energy block is a constant, i.e.,
\begin{align}\label{linear3}
\sum_{i=0}^{n-1}\Energy_i = 2\sum_{i=0}^{n-1}A_i^2 =2\sum_{a\in\mathcal{A}} a^2n_{a}.
\end{align}
In analogy to \eqref{linear2}, \eqref{linear3} shows that these $n$ symbol energies also satisfy a linear relationship. By taking expectation on both sides of \eqref{linear3}, we have
\begin{equation}\label{linear4}
\sum_{i=0}^{n-1}\Exp{E_i} = 2\sum_{a\in\mathcal{A}} a^2n_{a}.
\end{equation}

Subtracting \eqref{linear4} from \eqref{linear3} yields
\begin{align}\label{subsum}
    \sum_{i=0}^{n-1}(\Energy_i-\Exp{\Energy_i}) = 0,
\end{align}
which can be rewritten as
\begin{align}\label{subsum1}
    \Energy_j-\Exp{\Energy_j} = -\sum_{\substack{j'=0\\j'\neq j}}^{n-1}(\Energy_{j'}-\Exp{\Energy_{j'}}).
\end{align}

By using \eqref{EqDefAutoCov} and \eqref{subsum1}, $\Cov{\Energy_i,\Energy_{j}}$ is expanded, i.e.,
\begin{align}
  \Cov{\Energy_i,\Energy_{j}} =&  \Exp{\left(\Energy_i-\Exp{\Energy_i}\right)\left( -\sum_{\substack{j'=0\\j'\neq j}}^{n-1}(\Energy_{j'}-\Exp{\Energy_{j'}}) \right)} \\
  =&  -\Exp{\left(\Energy_i-\Exp{\Energy_i}\right)^2}  \nonumber \\
  & -\sum_{\substack{j'=0\\j'\neq j,i}}^{n-1}\Exp{\left(\Energy_i-\Exp{\Energy_i}\right)\left(\Energy_{j'}-\Exp{\Energy_{j'}}\right)}\\
  =&  -\Var{\Energy_i}-\sum_{\substack{j'=0\\j'\neq j,i}}^{n-1}\Cov{\Energy_i,\Energy_{j'}}.\label{covvar}
\end{align}

% Within the same block, the symbol energies at different time instants are interchangeable, as they follow the same probabilistic model. 
Therefore, \eqref{covvar} can be rewritten as 
\begin{align}
    %\Cov{\Energy_i,\Energy_{j}}+\sum_{\substack{j'=0\\j'\neq j,i}}^{n-1}\Cov{\Energy_i,\Energy_{j'}}=-\Var{\Energy_i}\label{covvar2}\\
    \Cov{\Energy_i,\Energy_{j}}+(n-2)\Cov{\Energy_i,\Energy_{j}}=-\Var{\Energy_i}\label{covvar1}.
\end{align}

By using \eqref{VarStat} in \eqref{covvar1}, the equality in \eqref{negCov} is obtained. The inequality in \eqref{negCov} clearly follows from the fact that the variance $\Var{|X|^2}$ is positive. The last step in the proof is to show that the autocorrelation $R_{\Energy}(i,\tau)=\CrS$ is given by \eqref{bitcorr0}. This follows from substituting \eqref{negCov} into \eqref{DefAutoCovStationary}. Because of the negative autocovariance, it can be observed that $\CrS$ is smaller than $\Exp{|X|^2}^2$. This completes the proof.

\section{Proof of Theorem~\ref{Thm:WndVar}}\label{Proof:WndVar}

To prove \eqref{CCTExp}, we use \eqref{wndEngDef} in \eqref{AvgExpWe} to obtain
\begin{align}
\AvgExp{\WE} &=
 \frac{1}{n}\sum_{i=0}^{n-1}\sum_{j=i-W/2}^{i+W/2} \Exp{|X_{j}|^2} \label{CCTExp1} \\ 
&=  \frac{1}{n}\sum_{i=0}^{n-1}\sum_{j=i-W/2}^{i+W/2} \Exp{|X|^2}\label{CCTExp0}\\
&=(W+1)\Exp{|X|^2},  
\end{align}
where \eqref{CCTExp1} uses the linearity of expectation, and \eqref{CCTExp0} follows from \eqref{EX2Stat}. %Therefore, \eqref{CCTExp} follows directly from \eqref{CCTExp0}.

To prove \eqref{wndTimeEngVar}, we begin with $\Var{\WE_i}$. %Due to the fact that $\Var{\WE_i}$ is the variance of the sum of $W+1$ symbol energies as shown in 
Using \eqref{wndEngDef}, Lemma~\ref{Thmmin1}, and \cite[Thm.~9.2]{yates2014probability}, we have
\begin{align}
\Var{\WE_i}=\; & \sum_{j=i-W/2}^{i+W/2}\Var{|X_j|^2} \nonumber \\
 & + 2\sum_{j=i-W/2}^{i+W/2-1}\sum_{\substack{j'=j+1}}^{i+W/2}\Cov{\Energy_{j},\Energy_{j'}} \label{wndEngVar0} \\
  =\; & (W+1)\:\Var{|X|^2} \nonumber \\
 & + 2\sum_{j=i-W/2}^{i+W/2-1}\sum_{\substack{j'=j+1}}^{i+W/2}\Cov{\Energy_{j},\Energy_{j'}} \label{wndEngVar1} \\
 = \; & (W+1)\:\Var{|X|^2} \nonumber \\
 & + 2\sum_{j=i-W/2}^{i+W/2-1}\sum_{\substack{j'=j+1}}^{i+W/2}\big[R_{\Energy}(j,j'-j) \nonumber \\
 & -\Exp{|X|^2}^2\big] \label{wndEngVar2}
 \\ 
 = \; & (W+1)\:\Var{|X|^2}-W(W+1)\Exp{|X|^2}^2 \nonumber \\
 & + 2\sum_{j=i-W/2}^{i+W/2-1}\sum_{\substack{j'=j+1}}^{i+W/2} R_{\Energy}(j,j'-j). \label{wndEngVar3}
\end{align}
where \eqref{wndEngVar1} follows from \eqref{VarStat}, and \eqref{wndEngVar2} from \eqref{DefAutoCovStationary}.

\begin{figure}[!t]
\centering
% \resizebox{0.6\linewidth}{!}{\input{./Figures/delay_tau.tikz}}
\includegraphics[width=0.6\linewidth]{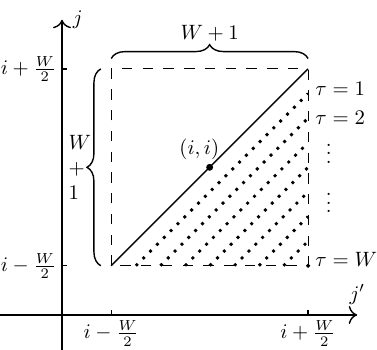}
\caption{An illustration of possible combinations of $j$ and $j'$ for the double summations of the last term in \eqref{wndEngVar3}. The dots at each diagonal direction represent $W-\tau+1$ possible pairs of $j$ and $j'$ when $j-j'>0$.}
\label{delay_tau}
\end{figure}

The first two terms in \eqref{wndEngVar3} are known, since $\Var{|X|^2}$ and $\Exp{|X|^2}$ are given in \eqref{VX2} and \eqref{EX2}. Meanwhile, $R_{\Energy}(j,j'-j)$ depends on the time instant $j$ and delay $\tau = j'-j>0$. Therefore, based on \eqref{AvgVarWe}, only the last term in \eqref{wndEngVar3} needs to be considered for averaging over $n$ time instants, i.e.,
\begin{align}
 & \frac{1}{n}\sum_{i=0}^{n-1}\left(2\sum_{j=i-W/2}^{i+W/2-1}\sum_{\substack{j'=j+1}}^{i+W/2} R_{\Energy}(j,j'-j)\right) \label{avgVar0}  \\
  =\; & 2\sum_{\tau=1}^{W}\frac{1}{n}\sum_{i=0}^{n-1}\sum_{j=i-W/2}^{i+W/2}R_{\Energy}(j,\tau)\label{avgVar1}\\
  =\; & 2\sum_{\tau=1}^{W}(W-\tau+1)\AC, \label{avgVar}
\end{align}
where \eqref{avgVar1} can be explained by observing Fig.~\ref{delay_tau}, which shows possible pairs of $j$ and $j'$ and the corresponding values of $\tau$. To obtain \eqref{avgVar}, we observe in Fig.~\ref{delay_tau} that for each value of $\tau$, there are $W-\tau+1$ pairs in the corresponding diagonal, whose $R_{\Energy}(j,\tau)$ are averaged. Hence, \eqref{avgVar} is obtained. Finally, $\AvgVar{\WE}$ in \eqref{wndTimeEngVar} is simply the sum of the first two terms in \eqref{wndEngVar3} and \eqref{avgVar}, which completes the proof.

\section{Proof of Theorem~\ref{Thm:LinEDI}}\label{Appx:LinEDI}
In this Appendix, we show that in the special case of $n\leq W+2$, the computation of $\Var{\WE_i}$ can be simplified by considering the symbol energy ``pattern" inside the window. This leads to a simpler expression for the EDI in \eqref{CCDMEDI_Lin}.

Given a window length $W$, the window has $W+1$ symbol energies. This window in general covers multiple symbol energy blocks, and thus we can write
\begin{equation}
    W+1=un+r,
\end{equation}
where $u\in \mathbb{N}$ is the maximum number of complete symbol energy blocks covered by the window, and $r$ is the remainder when $W+1$ is divided by $n$ ($0\leq r \leq n-1$).

\begin{figure}[!t]
\centering
% \resizebox{0.8\linewidth}{!}{\input{./Figures/sldWnd.tikz}}
\includegraphics[width=0.8\linewidth]{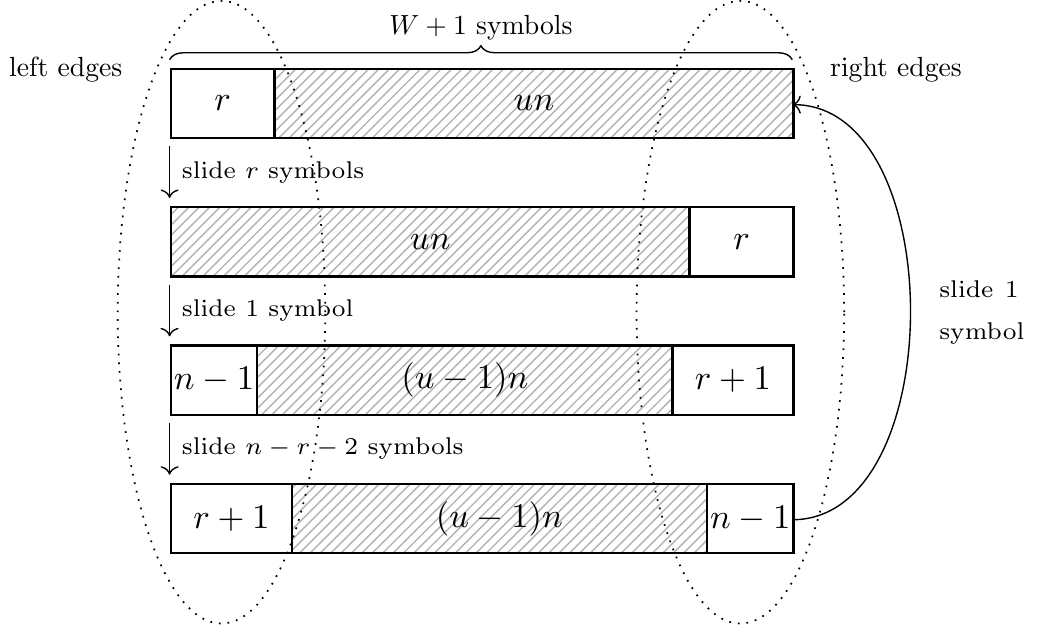}
\centering\caption{An illustration of four symbol energy patterns as window slides to the right through the symbol energy sequence. The shaded area represents $u$ complete blocks.}
\label{sldWnd}
\end{figure}

As the window slides, the symbol energy ``pattern" covered by the window varies cyclically with a period $n$, which is illustrated in Fig.~\ref{sldWnd}. The pattern shows that the sum of symbol energies with a shaded background is constant, whereas the energy of the incomplete blocks located at two edges of the windows is random. Therefore, the constant part of energy has no contribution to window energy variance, and thus can be removed in the computation of the variance. For example, for the pattern at the top of Fig.~\ref{sldWnd}, it consists of $u$ complete blocks at the right side, while at its left edge there are $r$ symbols from the adjacent block. Hence, the variance of the top pattern is simplified by only considering these $r$ symbols. Such simplification can by done by using \eqref{wndEngVar3}. After substituting $W+1$ with $r$, and $R_{\Energy}(j,j'-j)$ with $\CrS$, we have
\begin{align}
 \Var{\WE_i}=\ & r\Var{|X|^2}-r(r-1)\Exp{|X|^2}^2\nonumber \\
    &+r(r-1)\CrS. \label{cst1}
\end{align}
%With respect to \eqref{wndEngVar3}, \eqref{cst1}  is $\CrS$ as given in \eqref{bitcorr0}, since these in the $r$ symbols are mutually correlated.

As long as $W+1\geq n-1$, it can be concluded from Fig.~\ref{sldWnd} that: (i) two edges with random symbol energies contribute to the variance, and (ii) these $n$ patterns keep $0,1,...,n-1$ mutually correlated symbol energies at both edges, respectively. Under these circumstances, $\AvgVar{\WE}$ is obtained by averaging the variance over $n$ patterns, i.e.,    
\begin{align}
\AvgVar{\WE}  = &\frac{2}{n}\sum_{i=0}^{n-1}\bigg[i\Var{|X|^2}-i(i-1)\Exp{|X|^2}^2 \nonumber \\ 
 & +\:i(i-1)\CrS\bigg]\label{engPat1} \\
  = &\frac{2\Var{|X|^2}}{n(n-1)}\sum_{i=0}^{n-1}(ni-i^2)\label{engPat2} \\
  = & \frac{(n+1)\Var{|X|^2}}{3}\label{engPat},
\end{align}
where \eqref{engPat2} follows from \eqref{DefAutoCovStationary} and \eqref{negCov}.
%, and the summation in \eqref{engPat} is obtained from  \cite[Sec.~VI]{spiegel2009schaum}. 
By substituting \eqref{engPat} and \eqref{CCTExp} into \eqref{EDI}, EDI in \eqref{CCDMEDI_Lin} is obtained.

\section{Proof of Corollary~\ref{Thm:LowLmt}}\label{Appx:LowLmt}
We first discuss the bounds on $\AvgVar{\WE}$, which directly determine the bounds on EDI. For any finite blocklength $n$, the $\AvgVar{\WE}$ of the CCDM QAM symbol sequence satisfies
\begin{equation}\label{WndVarBounds}
   0 \leq \AvgVar{\WE} \leq (W+1)\:\Var{|X|^2}.
\end{equation}
The variance is always positive, hence the left inequality in \eqref{WndVarBounds} is true. Based on \eqref{wndEngVar1}, due to negative autocovariance from \eqref{negCov}, we have,
\begin{align}
 \Var{\WE_i} = & (W+1)\:\Var{|X|^2} \nonumber \\
  & + 2\sum_{j=i-W/2}^{i+W/2-1}\sum_{j'=j+1}^{i+W/2}\Cov{\Energy_{j},\Energy_{j'}}, \\
 \leq & (W+1)\Var{|X|^2}\label{ineq.W},
\end{align}
where \eqref{ineq.W} holds with equality when $W=0$. After using \eqref{AvgVarWe}, we have the right inequality in \eqref{WndVarBounds}. By using \eqref{WndVarBounds} and \eqref{CCTExp} in \eqref{EDI}, the inequalities for the EDI in \eqref{EDIBounds} is obtained.

We now prove how the bounds in \eqref{EDIBounds} are achieved asymptotically by $W$. In terms of the upper bound in \eqref{upLimit}, by setting $W = 0$, the window only encompasses one symbol energy, thereby $\AvgVar{\WE} = \Var{|X|^2}$ and $\AvgExp{\WE} = \Exp{|X|^2}$. With \eqref{EDI} and using \eqref{Kur}, \eqref{upLimit} is obtained. From \eqref{CCDMEDI_Lin}, it can be seen that for $n<W+2$, the EDI tends to $0$ for $W\rightarrow \infty$. This completes the proof. 

\ifCLASSOPTIONcaptionsoff
  \newpage
\fi

\bibliographystyle{IEEEtran}
\bibliography{NLIbib.bib}
\end{document}